\documentclass[a4paper,fleqn]{cas-dc}

\usepackage[numbers]{natbib}
\usepackage{graphbox}
\usepackage{tabularx}
\usepackage{amsmath}
\usepackage{subcaption}
\usepackage{multirow}
\usepackage{subcaption}
\usepackage{hyperref}
\usepackage{gensymb}
\usepackage{lipsum}
  {
      
  }

\def\tsc#1{\csdef{#1}{\textsc{\lowercase{#1}}\xspace}}
\tsc{WGM}
\tsc{QE}

\begin{document}
\let\WriteBookmarks\relax
\def\floatpagepagefraction{1}
\def\textpagefraction{.001}

\shorttitle{Wake effects in a 100 GW North Sea offshore wind scenario}    

\shortauthors{Quiñones et~al.}  

\title [mode = title]{Wake effects in a 100 GW North Sea offshore wind scenario: a model intercomparison
under year-round weather conditions} 



%

\author[1,2]{Jhon J. {Quiñones}}
\cormark[1]
\ead{jquinonescorte@tudelft.nl}

\author[3]{Diego Siguenza-Alvarado}
\author[4]{Peter Baas}
\author[4]{Pim {Van Dorp}}
\author[5]{Majid Bastankhah}
\author[6]{Remco A. Verzijlbergh}

\credit{<Credit authorship details>}


\affiliation[1]{organization={Faculty of Electrical Engineering, Mathematics and Computer Science, Delft University of Technology}, 
            city={Delft},
            postcode={2628 XE}, 
            country={the Netherlands}}  

\affiliation[2]{organization={School of Mechanical Engineering, Purdue University}, 
            city={West Lafayette},
            postcode={47907}, 
            state={IN},
            country={USA}}

\affiliation[3]{organization={Centro de Energías Renovables y Alternativas, Escuela Superior Politécnica del Litoral, ESPOL, Campus Gustavo Galindo km. 30.5 Vía Perimetral}, 
            city={Guayaquil},
            postcode={090902}, 
            country={Ecuador}}

\affiliation[4]{organization={Whiffle BV}, 
            city={Delft},
            postcode={2629 JD}, 
            country={the Netherlands}}

\affiliation[5]{organization={Department of Engineering, University of Durham}, 
            city={Durham},
            postcode={DH1 3LE}, 
            country={UK}}

\affiliation[6]{organization={Faculty of Technology, Policy and Management, Delft University of Technology}, 
            city={Delft},
            postcode={2628 BX}, 
            country={the Netherlands}}            
            
\credit{}

\cortext[1]{Corresponding author}

\begin{abstract}
Large-scale offshore wind development will intensify wake interactions between wind farm clusters in the North Sea, increasing uncertainty in regional energy estimates. We assess HARMONIE--AROME mesoscale simulations, ASPIRE large-eddy simulations (LES), and three engineering wake models based on Jensen, Bastankhah Gaussian, and TurbOPark formulations for a 100.7~GW 2050 offshore wind scenario. We find that LES and mesoscale simulations accurately represent observed local free-stream wind conditions. Moreover, all engineering wake models predict higher mean wind speeds than LES in every month, with biases reaching 0.60–0.70~m~s$^{-1}$ at 10–40~km downstream. These differences also vary systematically with atmospheric regime: Bastankhah Gaussian model remains positively biased across wind speed and stability conditions, whereas Jensen and TurbOPark shift from underprediction in unstable and near-neutral conditions to overprediction in stable conditions at 8–12~m~s$^{-1}$. These results show that agreement in annual capacity factor can conceal substantial differences across space, time, and atmospheric regime, highlighting the importance of including sensitivity to model formulation in regional wind energy assessments.
\vspace{0.5cm}
\end{abstract}

\maketitle


\section{Introduction}


Offshore wind is expected to play a central role in Europe’s transition toward a climate-neutral and secure energy system \cite{Wiese2024,Lal2026}. The nine countries participating in the North Seas Energy Cooperation have committed to developing at least 120~GW of offshore wind capacity by 2030 and 300~GW by 2050 \cite{EuropeanCommissionNSEC,hamburg2026_declaration}. Meeting these targets will require a transition from largely separated projects to clusters of multi-gigawatt wind farms. This expansion will place more turbines within the same regional atmospheric flow, allowing wakes to extend beyond project boundaries and reduce the wind resource available to neighboring farms. These wider interactions are expected to make wake induced \textit{aerodynamic losses} more extensive and more consequential for regional energy production than those observed today \cite{Akhtar2021,MaasRaasch2022,Borgers2024}. Quantifying these losses under future deployment scenarios is essential for assessing the energy production associated with the planned North Sea capacity.


Wind turbine wakes develop as turbines extract momentum and kinetic energy from the atmospheric flow. The resulting wind speed deficits and enhanced turbulence persist downstream, reducing the power available to subsequent turbines \cite{StevensMeneveau2017,PorteAgel2020}. Within individual wind farms, repeated exposure to these deficits produces internal wake losses. As wind farms become larger and more closely spaced, these effects can extend beyond individual farms and develop into cluster scale wakes. Evidence of such interactions is already available from operating offshore clusters. Measurements, remote sensing, and production data from the German North Sea have identified wake related power reductions at a wind farm located up to 55~km downstream of the generating farms \cite{Schneemann2020}. Regional simulations indicate that these effects could strengthen as offshore development expands. Annual mean wind speed deficits of 2--2.5~m~s$^{-1}$ have been projected within wind farm areas, with average wake extents of approximately 40--45~km and capacity factor reductions exceeding 20\% at strongly exposed locations \cite{Akhtar2021}. Under shallow boundary layer conditions, large-eddy simulations indicate that wakes from very large wind farms can persist for more than 100~km \cite{MaasRaasch2022}. At these larger spatial scales, wake effects can also involve changes in the capping inversion, pressure gradients, wind direction, and atmospheric gravity waves \cite{AllaertsMeyers2017,MaasRaasch2022,LanzilaoMeyers2024}. The strength and recovery of these effects vary with wind speed, wind direction, atmospheric stability, boundary layer depth, and deployment geometry \cite{Schneemann2020,MaasRaasch2022}. Representing this variability across realistic annual weather conditions is a central challenge for regional wind resource and energy yield assessment.


Practical wind resource and energy yield assessments rely on models that differ widely in physical detail and computational cost \cite{Gocmen2016,PorteAgel2020,Ouro2025}. These include analytical engineering wake models, mesoscale atmospheric models with wind farm parameterizations, and atmospheric large-eddy simulation. Among these approaches, analytical wake models remain the principal tools for wind farm design, layout optimization, scenario screening, and annual energy production assessment. Their low computational cost allows large numbers of turbines, atmospheric conditions, and layouts to be evaluated efficiently \cite{Gocmen2016,PorteAgel2020}. Foundational approaches include the Jensen model, which represents the wake as a linearly expanding top-hat velocity deficit \cite{Jensen1983}, and the Gaussian model of Bastankhah and Porté-Agel, which is based on momentum conservation \cite{BastankhahPorteAgel2014}. Modern engineering frameworks build on these approaches by including models for wake expansion, added turbulence, deflection, rotor averaging, and the superposition of overlapping wakes \cite{PorteAgel2020,Ouro2025}. This reduced representation makes annual and regional calculations feasible, but requires wake development to be described through empirical or semiempirical relationships.

The use of analytical models has expanded from isolated turbine wakes to complete offshore farms and interacting clusters. Comparisons with operational data from Horns Rev, Lillgrund, and Anholt showed that sector averaging, wind direction uncertainty, background gradients, and model calibration can substantially affect predicted array efficiency \cite{Gaumond2012,Pena2018}. More recent comparisons with lidar and SCADA data show that these models can reproduce important features of mean wake deficits and power production. However, their predictions remain sensitive to wake growth, turbulence, superposition, and calibration choices \cite{Zhan2020,VanBinsbergen2024,SiguenzaAlvarado2026}. As offshore developments have increased in size, model development has also focused on the longer wakes and deeper arrays associated with large clusters. TurbOPark was introduced to improve wake recovery in large arrays and over long downstream distances \cite{Nygaard2020,Pedersen2022}. A benchmark involving 19 offshore wind farms reported improved performance relative to traditional Park formulations \cite{Nygaard2022}. These developments have enabled engineering models to assess annual wake interactions between neighboring North Sea clusters, but predicted losses remain sensitive to model formulation and can differ substantially from mesoscale estimates \cite{Porchetta2024}. Although computationally efficient, these models require careful evaluation for future regional clusters, where wake recovery and atmospheric replenishment extend beyond the scales of most existing wind farms. 


Mesoscale atmospheric models provide a second approach to assess wind resource and energy yield. These models simulate regional weather over domains of hundreds of kilometers and periods from months to decades, capturing synoptic forcing, coastal gradients, atmospheric stability, and boundary layer evolution. Because their kilometer scale grids do not resolve individual turbines or wakes, wind farm effects are represented through parameterizations. Common schemes either distribute turbine momentum extraction through the rotor layer or describe wake expansion within a grid cell. They differ in their treatment of extracted energy, turbulence, and wake structure \cite{Fitch2012,Volker2015,Ouro2025}.Comparisons of the Weather Research and Forecasting (WRF) simulations with aircraft measurements over wind farms in the German Bight showed that vertical turbulent momentum transport controlled wake recovery under stable conditions. The results also depended on grid spacing and on whether turbulence generated by turbines was included \cite{Siedersleben2020}. Five WRF wind farm parameterizations applied to a North Sea case likewise produced substantial differences in wind speed deficit, turbulence, and estimated power \cite{Ali2023}. Despite this sensitivity, mesoscale models have enabled annual and regional assessments of large offshore systems. Simulations of very large developments estimated that wakes within farms and from neighboring farms could reduce total production by about one third \cite{Pryor2021}. Experiments with COSMO--CLM also showed that increasing capacity density from 3.5 to 10~MW~km$^{-2}$ reduced the integrated annual capacity factor from 51.8\% to 38.2\%. The magnitude depended on turbine size, farm spacing, and the number of upstream projects \cite{Borgers2024}. In a 92~GW North Sea expansion scenario, 25 of the 69 wind farms operating in 2024 could lose more than 5\% of their annual production, with losses exceeding 10\% at 13 farms \cite{BorgersMeyersVanLipzig2025}. \citet{Dorenkamper2025} similarly projected that annual capacity factors in Germany and the Netherlands could decline from approximately 47\% in 2025 to 36\% and 43\%, respectively, by 2050. These studies demonstrate the value of mesoscale models for regional assessment, while estimates of internal wakes and power remain sensitive to spatial resolution and parameterization.


LES provides a more detailed representation of atmospheric turbulence than mesoscale models. Mesoscale models parameterize processes at the turbine scale, whereas LES resolves the larger turbulent motions that transport momentum and heat within the atmospheric boundary layer. Motions smaller than the numerical grid are represented through a subgrid model \cite{Mehta2014,StevensMeneveau2017,PorteAgel2020} and turbines are commonly represented using actuator disk or actuator line methods. This level of detail allows LES to describe turbine generated turbulence, wake merging, lateral and vertical momentum transport, and the evolving interaction between a wind farm and the overlying atmosphere \cite{StevensMeneveau2017,PorteAgel2020}. This capability has made LES an important tool for investigating the physics of large wind farms \cite{AllaertsMeyers2017,LanzilaoMeyers2024}. Recent studies have extended LES from investigations of physical processes to practical wind resource and energy yield assessment. \citet{Baas2023} simulated one year of actual weather for several hypothetical 4~GW offshore wind farms and quantified the sensitivity of annual production to turbine rating, capacity density, layout, stability, and global blockage. \citet{Postema2025} used a simulation spanning ten years to assess how shorter simulations can support estimates of long term annual energy production. Comparisons with operational data have also shown promise. Using two years of SCADA measurements from Hollandse Kust Noord, \citet{DeJong2026} reported an annual energy error below 1\% for the ASPIRE LES configuration and smaller wake loss errors than the steady state engineering models included in the comparison. However, most LES applications still focus on individual farms, idealized clusters, or limited regional subsets. Simulations spanning a full year over large domains containing many interacting clusters remain computationally demanding and rare.


Across the three modeling approaches, studies of future deployment scenarios generally project increasing wake losses as installed capacity and deployment density rise. However, substantial uncertainty remains about the magnitude of the resulting energy penalty, particularly for wind farms and clusters approaching the scale planned for the North Sea. Recent work on theoretical limits of offshore wind energy extraction further shows that regional production estimates depend on how models represent atmospheric momentum replenishment and finite farm effects \cite{Ferreira2025,VanDerLaan2026}.

This dependence on model representation has motivated direct intercomparison studies involving different combinations of modeling approaches. \citet{Fischereit2022} compared engineering, Reynolds averaged, and mesoscale models and found differences in the representation of internal wakes, interactions between farms, heterogeneous background flow, and downstream recovery. Other studies have compared engineering models with LES. \citet{Centurelli2021} found that analytical induction superposition underestimated the upstream velocity deficit relative to LES as farm size increased and boundary layer height decreased. Similarly, the 15~GW wind farm study conducted by \citet{Maas2023} linked substantial power differences between analytical wake models and LES to atmospheric pressure work, geostrophic forcing, and energy dissipation that are not represented in conventional wake models. These differences are not uniform across spatial scales or atmospheric conditions. In a full year comparison of neighboring North Sea clusters, \citet{Porchetta2024} found substantial differences between WRF and several engineering wake models, particularly in the magnitude and spatial distribution of external wakes. Newer engineering formulations reproduced the mesoscale wake pattern more consistently than traditional Jensen and Gaussian implementations, although the predicted magnitudes still varied widely \cite{Porchetta2024}. Comparisons between mesoscale models and LES reveal another aspect of this variability. For three planned wind farms on the United States East Coast, \citet{SanchezGomez2026} found that mesoscale simulations captured the broader cluster wake behavior and its dependence on atmospheric stability. However, they underestimated internal wake power losses, especially for aligned turbines and stable conditions.

Despite this growing body of comparison studies, most include only two of the three principal modeling approaches and cover limited domains or periods. Evaluating all three approaches over a full year for the same hypothetical deployment and meteorological period can establish which estimates of regional wake loss and energy production remain consistent across models. It can also reveal how the remaining differences vary over time, with distance from wind farms, and across atmospheric conditions.


The WINS50 dataset provides the basis for this evaluation. It includes ASPIRE LES output for the 2020 and 2050 wind farm deployments, together with several HARMONIE--AROME simulations \cite{Baas2024_wins50}. The HARMONIE--AROME dataset covers 2019--2021 for both the actual wind farm deployment and corresponding free stream conditions without wind farms, as well as the 2050 deployment used in the regional comparison. The comparison assessment presented in this study focuses on the meteorological conditions of 2020. Figure~\ref{fig:domain_overview} provides the spatial context for the comparison. It shows the 2020 and 2050 wind farm layouts, the 373 atmospheric evaluation points, the 31 virtual masts used for the stability analysis, the observational sites, and illustrative Points~A and B. Differences in model forcing, spatial resolution, turbine representation, and available outputs are considered when interpreting the broader comparison. 

To our knowledge, this is the first full year regional study to compare engineering wake models, mesoscale atmospheric modeling, and LES for a common hypothetical offshore deployment. The analysis comprises two complementary comparisons. First, after evaluating HARMONIE--AROME and LES against offshore observations, we compare the regional wind and power fields produced by the three modeling approaches. Second, we conduct a more controlled comparison in which configurations based on Jensen (Model~A), Bastankhah Gaussian (Model~B), and TurbOPark (Model~C) use common LES background fields and turbine definitions, with LES serving as a higher fidelity numerical reference. We quantify monthly differences in wind speed and turbine power and examine how they vary with distance from wind farms, wind speed, and atmospheric stability.

\section{Methodology}
\label{sec:methods}

This study compares three modeling approaches for the WINS50 offshore wind energy deployment scenario: \textit{(i)} HARMONIE-AROME mesoscale simulations, \textit{(ii)} ASPIRE large-eddy simulations (LES), and \textit{(iii)} three engineering wake model configurations implemented in \textsc{FOXES}. The approaches resolve different processes and are not expected to agree pointwise. HARMONIE-AROME represents regional atmospheric adjustment and wind farm effects at kilometer-scale resolution, ASPIRE resolves a substantial fraction of the turbulent flow and represents individual turbines as actuator disks, and the engineering models apply reduced-order wake relations to prescribed background fields. All analyses refer to the meteorological year 2020. This means that the simulations represent the same calendar year; it does not imply identical atmospheric forcing, because the mesoscale and LES frameworks have distinct forcing, nesting, and data-assimilation procedures.

The comparison has two parts. First, wind speeds from HARMONIE-AROME and ASPIRE are evaluated against observations and their 2050 wind and power fields are compared descriptively. Second, the three engineering configurations are evaluated against the 2050 LES in a controlled comparison. In this second part, all engineering configurations use the same LES-derived background fields and turbine metadata. The LES is treated as a higher-fidelity numerical reference rather than as ground truth. Consequently, an engineering-model-LES difference is a composite
model discrepancy and cannot, by itself, be attributed uniquely to one physical process.

\subsection{Wind farm scenarios and analysis locations}
\label{sec:scenarios_locations}

In this study, two wind energy deployment scenarios were considered. The 2020 scenario represents the operational turbine fleet and comprises 3,470 turbines with a total installed capacity of 12.8~GW. The hypothetical 2050 scenario was constructed based on the 2050 offshore wind energy targets of the North Sea countries as they stood in 2018, assuming a capacity density of 8~MW~km$^{-2}$ and a maximum wind farm size of 400~km$^{2}$ \cite{Baas2024_wins50}. This scenario adds 6,593 turbines, resulting in 10,063 turbines and a total installed capacity of 100.7~GW. Offshore layouts, turbine types, hub heights, rotor diameters, power curves, and thrust curves were harmonized across the model set wherever the model formulation allowed. Onshore turbine locations and hub heights were obtained from the
\textit{Windturbines - ashoogte} dataset \cite{AtlasLeefomgeving2026}.

Figure~\ref{fig:domain_overview} shows the LES and wake engineering models domain, the turbine locations for the 2020 and 2050 scenarios, and the 373 atmospheric evaluation points. The latter comprise all grey, green and red markers, together with the four individually highlighted locations. The green markers identify the subset of 31 virtual masts used for the stability analysis, while distinct symbols identify the K13A and Europlatform sites and points A and B used for the illustrative time series.

\subsection{Mesoscale simulation}
\label{sec:mesoscale}

The mesoscale simulations were performed with the HARMONIE--AROME numerical weather prediction model, which is used operationally by several European meteorological services, including KNMI \cite{Bengtsson2017}. HARMONIE--AROME is a limited-area, non-hydrostatic model that explicitly solves the vertical momentum equation and represents atmospheric processes relevant to offshore wind conditions. The WINS50 configuration used HARMONIE--AROME Cycle~43, together with the HARATU turbulence scheme \cite{DeRooy2022} and the SURFEX~7.3 land-surface model. A detailed description of the model configuration is provided by \citet{vanStratum2022}.

The simulation domain followed the configuration of the Dutch Offshore Wind Atlas and comprised $789\times789$ grid points with a horizontal grid spacing of 2.5~km. The domain was centered near Cabauw and covered the North Sea and the surrounding land areas. Lateral boundary conditions and sea-surface temperatures were prescribed from ERA5 using the Davies--Kallberg relaxation scheme \cite{Davies1976,Hersbach2020}. In addition to the observations assimilated in ERA5, conventional observations, Mode-S EHS aircraft measurements, and ASCAT scatterometer winds were assimilated using a three-dimensional variational scheme \cite{DeHaan2011,DeHaan2016,MarseilleStoffelen2017}.

Two mesoscale configurations were used in the present study. The first was a reference simulation without wind farm effects. This simulation extended the DOWA climatology, originally covering 2008--2018, with the years 2019--2021 \cite{Wagenaar2020} and provided the undisturbed atmospheric conditions against which wind-farm-induced changes were evaluated. DOWA was produced with Cycle~40, whereas the WINS50 extension used Cycle~43. The effect of this model-cycle transition was assessed by \cite{Theeuwes2021}. Differences in offshore wind speed at 100~m remained below 2\%, supporting the consistency of the combined climatology.

The second configuration included wind farm effects through the Fitch wind farm parameterization \cite{Fitch2012}. Within this parameterization, each turbine in a grid cell applies a momentum sink over the model levels intersecting the rotor and contributes to the diagnosed electrical power according to its power and thrust curves. Part of the extracted kinetic energy is introduced as turbulent kinetic energy, allowing the aggregate influence of the wind farms on the resolved atmospheric flow to be represented. The wind farm layouts and installed capacities followed the scenarios described in the Appendix \ref{app}.

\subsection{Large-eddy simulations}
\label{sec:les}

The large-eddy simulations were performed with the ASPIRE model. This model solves the Navier--Stokes equations under the Boussinesq approximation, together with conservation equations for heat and moisture and representations of sub-grid turbulence, radiation, thermodynamics, microphysics, and wind turbine forcing. In this approach, the equations are solved on a numerical grid sufficiently fine to capture a large part of the turbulent spectrum. A concise description of the governing equations is provided below, while further details are given by \citet{Baas2023}.

The continuity equation is given by:
\begin{equation}
    \frac{\partial \rho_0 u_j}{\partial x_i}=0,
    \label{eq:les_continuity}
\end{equation}
where $\rho_0(z)$ is a base-state density profile.

The Navier--Stokes equation solved in the ASPIRE LES is given by
\begin{equation}
    \frac{\rho_0\partial u_i}{\partial t}
    =
    F_i
    -\frac{\partial \tilde{p}}{\partial x_i}
    -\frac{\partial p^h}{\partial x_i}
    -\rho g\delta_{i3},
    \label{eq:les_momentum}
\end{equation}
where $F_i$ contains several forcing terms, including the sub-grid contributions and turbine forces. The pressure is separated into a hydrostatic component and fluctuations around it:
\begin{equation}
    p=p^h+\tilde{p}.
    \label{eq:les_pressure}
\end{equation}

The hydrostatic-pressure-gradient term is given by
\begin{equation}
    \frac{\partial p^h}{\partial x}
    =
    p^h\frac{g}{R_d}
    \left(
    \int_z^{z_{\mathrm{top}}}
    \frac{\partial}{\partial x}
    \frac{1}{T_v(x,y,z)}
    \,\mathrm{d}z
    \right).
    \label{eq:les_hydrostatic}
\end{equation}

In addition to the momentum equations, ASPIRE solves conservation equations for the heat and moisture scalar variables:
\begin{equation}
    \rho\frac{\partial \phi}{\partial t}
    =
    S_\phi
    -\frac{\partial \rho u_j\phi}{\partial x_j}
    -\frac{\partial F_j^\phi}{\partial x_j}.
    \label{eq:les_scalars}
\end{equation}
Here, $\phi\in\{q_t,\vartheta_l\}$ denotes the total specific humidity or a temperature variable based on moist static energy. The total specific humidity is the sum of water vapour, liquid water, and ice:
\begin{equation}
    q_t=q_v+q_l+q_i.
    \label{eq:les_total_humidity}
\end{equation}

The temperature variable $\vartheta_l$ is based on the moist static energy $h_l$:
\begin{equation}
    \vartheta_l=\frac{h_l}{c_p},
    \label{eq:les_temperature}
\end{equation}
with
\begin{equation}
    h_l=c_pT+gz-L_vq_l-L_iq_i.
    \label{eq:les_moist_static_energy}
\end{equation}

These thermodynamic variables enter the hydrostatic-pressure calculation through the virtual temperature. The virtual temperature used in Equation~\ref{eq:les_hydrostatic} is given by:
\begin{equation}
    T_v
    =
    T\left[
    \frac{q_v}{\epsilon}+(1-q_v)
    \right].
    \label{eq:les_virtual_temperature}
\end{equation}

This formulation allows open boundary conditions and differs from the configuration described by \citet{Baas2023}, where periodic boundary conditions were used.

Wind turbines were represented as actuator disks. The turbine force, which forms part of $F_i$ in Equation~\ref{eq:les_momentum}, was calculated as:
\begin{equation}
    F_t
    =
    -\frac{1}{2}\rho A C_t^{\prime}\overline{M_d}^{\,2},
    \label{eq:les_turbine_force}
\end{equation}
and the turbine power was calculated as:
\begin{equation}
    P_t
    =
    \frac{1}{2}\rho A C_p^{\prime}\overline{M_d}^{\,3}.
    \label{eq:les_turbine_power}
\end{equation}

The $C_t^{\prime}$ and $C_p^{\prime}$ curves were generated in a separate preprocessing step. The manufacturer power and thrust curves are functions of the 10-min average free-stream wind speed and are valid under specified conditions. These curves were mapped to the required $C_t^{\prime}$ and $C_p^{\prime}$ curves, which are functions of the rotor-disk-averaged wind speed $M_D$.

Year-long LES simulations were performed for the two wind energy scenarios. Both simulations used the same meteorological conditions from 2020, allowing the effect of increased wind energy deployment to be isolated.

The LES domain covers 327.6 $\times$ 491.5 $\mathrm{km}^2$ and is centered near 53.34 $^\circ$N, 4.12$^\circ$E. A horizontal grid spacing of 128~m was used, resulting in 2560 $\times$ 3840 horizontal grid cells. The domain extends to an altitude of 6~km and contains 64 vertical levels. The lowest model layer is 32~m thick, while vertical grid stretching is applied above 340~m. Sub-grid-scale turbulence is represented through the LES closure model. To bridge the resolution gap between ERA5 and the LES grid, an intermediate mesoscale simulation was used. This outer domain was directly forced by ERA5 and provided the boundary conditions for the LES domain. The mesoscale domain spans 768~km, has a horizontal grid spacing of 3~km, and extends to an altitude of 10~km.

The LES runs were performed as daily simulations. Each day was initialized from the final three-dimensional state of the previous day, making the annual simulation effectively continuous. For computational purposes, the two annual simulations were divided into four half-year periods, each preceded by one day of spin-up. The simulations generated high-resolution atmospheric fields and selected output at the analysis locations. The computations were carried out on an AWS high-performance cloud cluster using four P4d instances, each equipped with eight NVIDIA A100 GPUs. The ASPIRE WINS50 simulation data are publicly available through the Registry of Open Data on AWS \cite{WhiffleWINS50OpenData}.

\subsection{Engineering wake-model simulations}
\label{sec:engineering_wake_models}

The engineering-model calculations were performed with the open-source Farm Optimization and eXtended yield Evaluation Software
(\textsc{FOXES}) \cite{Schmidt2023}. In contrast to the LES, these calculations do not solve the three-dimensional atmospheric flow. Instead, they combine reduced-order descriptions of the velocity deficit and wake-added turbulence generated by individual turbines with model-specific superposition rules. Three configurations were considered, denoted Models~A--C. Model~A uses the Jensen top-hat velocity-deficit model \cite{jensen1983note}, Model~B uses the axisymmetric Gaussian model of \citet{bastankhah2014new} together with the Crespo--Hern\'andez wake-added-turbulence formulation
\cite{crespo1996turbulence}, with its far-wake coefficients replaced by the turbulence-optimised values of \citet{zehtabiyan2024wind}, and Model~C uses the TurbOPark velocity-deficit formulation \cite{pedersen2022turbulence} together with the IEC/Frandsen wake-added-turbulence model
\cite{frandsen2007turbulence,IEC61400-1_2019}. Their principal
differences are summarised in Table~\ref{tab:model_summary}, and their complete configurations and formulations are documented in
Appendix \ref{app}.

All three configurations were forced by the spatially and temporally varying wind speed, wind direction and turbulence intensity extracted from the LES simulation of the 2020 reference scenario. The ambient wind speed and turbulence-intensity fields are denoted by $U_0(\boldsymbol{x},t)$ and $\mathrm{TI}_0(\boldsymbol{x},t)$, respectively, where $\boldsymbol{x}$ is spatial position and $t$ is time. Together with the corresponding wind-direction field, these fields already contain the flow effects of the operational 2020 turbine fleet. The engineering-model calculations represented only the additional turbines in the 2050 scenario; the operational turbines were not modelled again, thereby avoiding double counting of their wakes. Throughout this subsection, $\mathrm{TI}$ denotes turbulence intensity and $\mathrm{TI}_{\mathrm{eff}}$ its local wake-corrected value.

The calculations used the \texttt{Iterative} algorithm in
\textsc{FOXES}. Wake trajectories were represented by two-dimensional streamlines following the spatially varying LES-derived horizontal wind-direction field, and rotor-effective quantities were evaluated using a nine-point rotor grid. No model parameter was calibrated against the present LES results. All model choices were specified independently of these results, either through the executed \textsc{FOXES} configurations or, for the modified far-wake coefficients of the Model~B added-turbulence formulation, directly in the installed source. The exact model selections and numerical settings are provided in Appendix \ref{app}.

The wake-length treatments differed among the configurations. Models~A and C used a maximum wake length of 20~km, whereas Model~B imposed no explicit maximum wake length. This difference resulted from the version-specific default settings used in the respective \textsc{FOXES} calculations and is retained here because of its potential relevance to the distance-dependent comparison. The engineering-model atmospheric variables were evaluated at 373
points at a height of 100~m; turbine-level outputs were evaluated at the locations of the 6,593 turbines added in the 2050 scenario. The three engineering wake-model configurations are summarized in
Table~\ref{tab:model_summary}.

\begin{table*}[t]
\centering
\caption{Principal formulations and prescribed parameters of the three
engineering wake-model configurations.}
\label{tab:model_summary}

\begin{tabularx}{\textwidth}{
>{\raggedright\arraybackslash}p{0.21\textwidth}
>{\raggedright\arraybackslash}X
>{\raggedright\arraybackslash}X
>{\raggedright\arraybackslash}X}
\toprule
Component
& Model A (Jensen-based)
& Model B (Bastankhah-based)
& Model C (TurbOPark-based) \\
\midrule

Velocity-deficit profile
& Top-hat
& Gaussian
& Gaussian \\

Induction model
& Betz
& Madsen
& Madsen \\

Wake-growth parameter, $k_j$
& $0.04$
& $0.35\,\mathrm{TI}_{\mathrm{eff},j}$
& $0.04\,\mathrm{TI}_{\mathrm{eff},j}$ \\

Velocity-deficit superposition
& Linear
& Linear
& Local quadratic \\

Wake-added turbulence model
& None
& Crespo--Hern\'andez with Zehtabiyan-Rezaie far-wake coefficients
& IEC/Frandsen formulation \\

Turbulence-intensity superposition
& Not applicable
& Maximum wake-added contribution combined quadratically with
  $\mathrm{TI}_0$
& Maximum wake-added contribution combined quadratically with
  $\mathrm{TI}_0$ \\

Maximum wake length
& 20~km
& No explicit maximum
& 20~km \\

\bottomrule
\end{tabularx}

\begin{minipage}{\textwidth}
\footnotesize
\textit{Note.}
TI denotes turbulence intensity. The subscript $j$ identifies the
wake-generating turbine, and $k_j$ is its wake-growth parameter.
$\mathrm{TI}_{\mathrm{eff},j}$ is the local wake-corrected turbulence
intensity at turbine $j$, whereas $\mathrm{TI}_0$ is the ambient
turbulence intensity. IEC denotes the International Electrotechnical
Commission.
\end{minipage}
\end{table*}

\section{Results and discussion}
\label{sec:results}

\begin{figure*}
    \centering
    \includegraphics[width=0.7\linewidth]{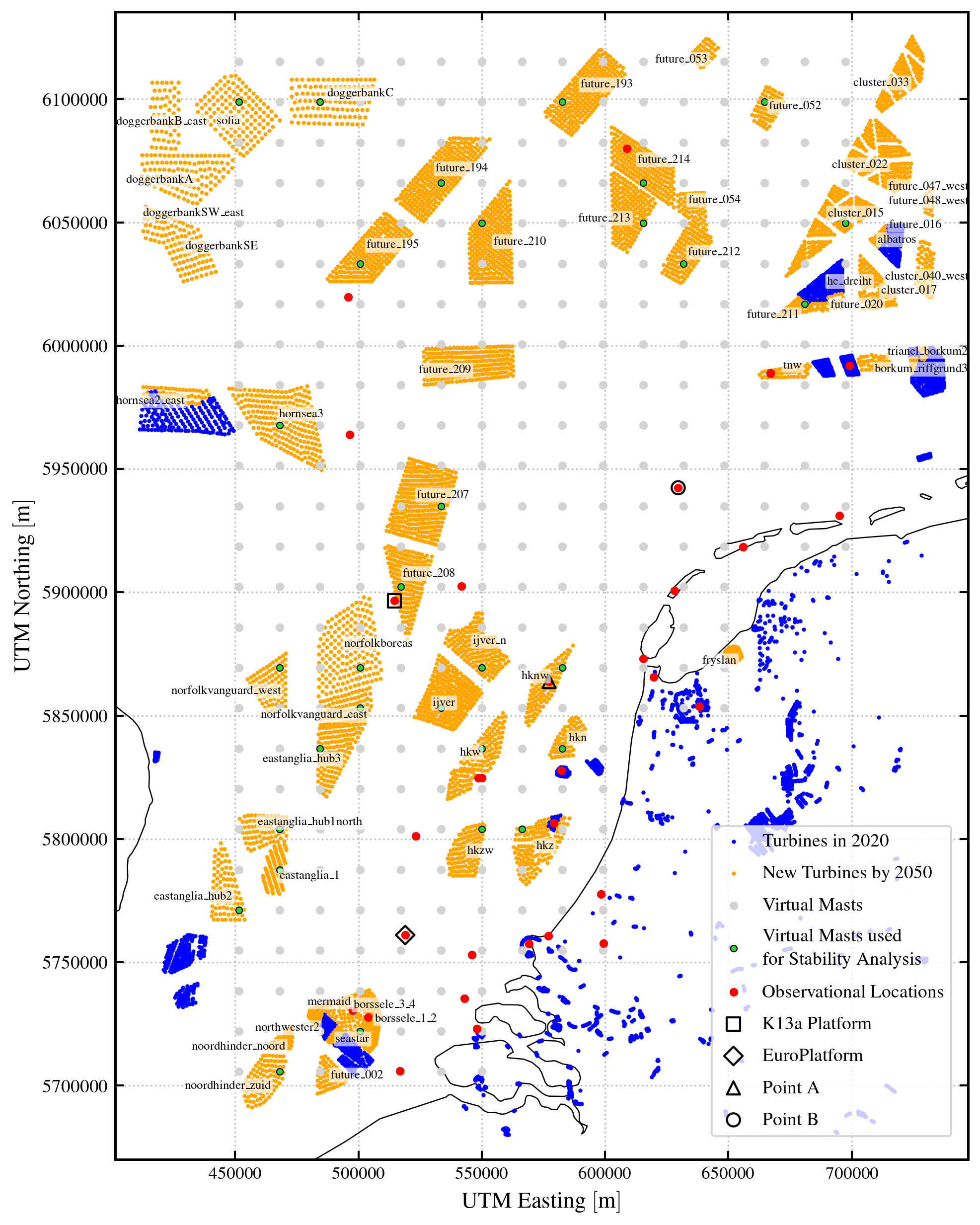}
    \caption{Overview of the LES and engineering-model domain. Blue and orange markers show the turbines in the 2020 operational fleet and the additional turbines in the 2050 scenario, respectively. The 373 atmospheric evaluation points comprise all grey, green and red markers and the four individually highlighted points. Grey markers denote virtual masts, green markers identify the subset of 31 virtual masts used for the stability analysis, and red markers denote observational locations. The square, diamond, triangle and circle identify K13A, Europlatform, point A and point B, respectively.}
    \label{fig:domain_overview}
\end{figure*}

\subsection{Validation and illustrative events}

Both the LES and mesoscale models have been run for the reference year 2020, including all the wind farms that were operational at the time. This allows for a validation against met-mast and floating LiDAR observations that covered the same simulation period. 

\begin{figure*}
    \centering
    \begin{subfigure}{0.49\textwidth}
        \includegraphics[width=\textwidth]{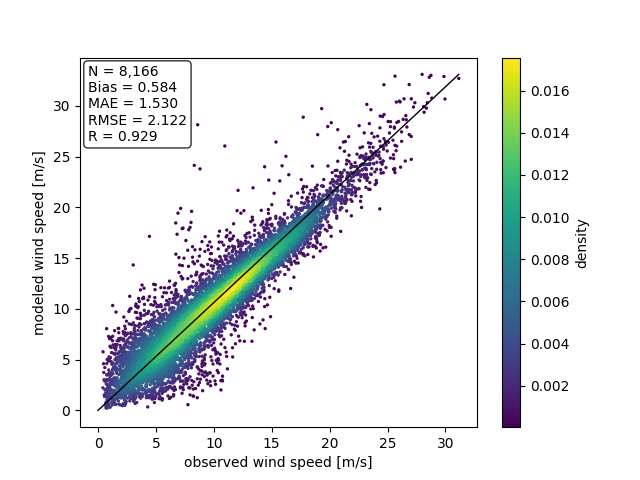}
        \caption{ASPIRE Europlatform}
        \label{fig:scatter-EPL-ASPIRE}
    \end{subfigure}
    \hfill
    \begin{subfigure}{0.49\textwidth}
        \includegraphics[width=\textwidth]{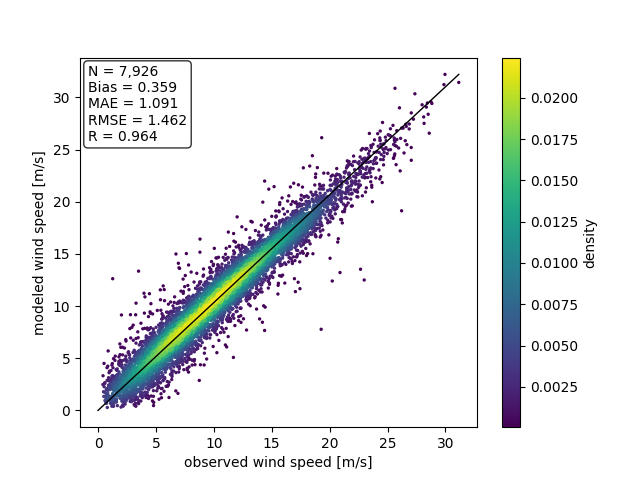}
        \caption{HARMONIE Europlatform}
        \label{fig:scatter-EPL-HARMONIE}
    \end{subfigure}
    \begin{subfigure}{0.49\textwidth}
        \includegraphics[width=\textwidth]{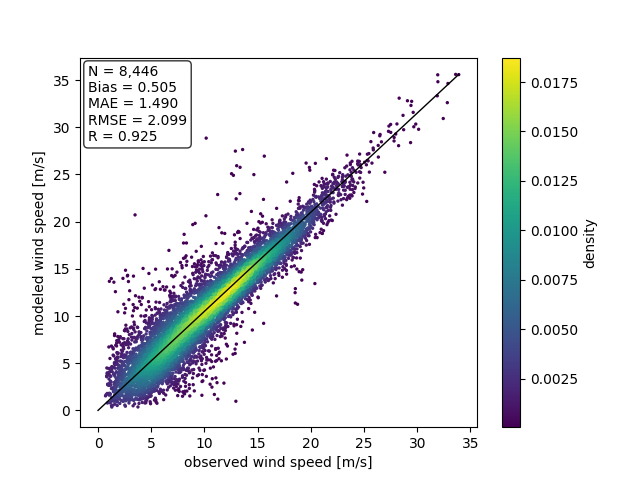}
        \caption{ASPIRE K13a}
        \label{fig:scatter-K13-ASPIRE}
    \end{subfigure}
    \hfill
    \begin{subfigure}{0.49\textwidth}
        \includegraphics[width=\textwidth]{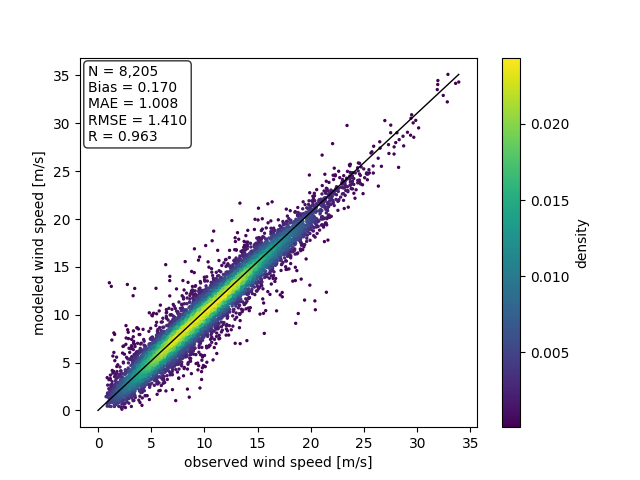}
        \caption{HARMONIE K13a}
        \label{fig:scatter-K13-HARMONIE}
    \end{subfigure}

    \caption{\label{fig:scatter-validation} Scatter plots of wind speed against North-Sea observations at the EuroPlatform and the K13A site.}
    \label{fig:figures}

\end{figure*}

Figure \ref{fig:scatter-validation} shows scatter plots of the LES and HARMONIE wind speeds against two observation sites, the Europlatform and K13A platform. The scatter plots show that both models capture the wind climate well, but the HARMONIE data shows better error statistics, both in terms of scatter (measured by MAE, RMSE and correlation coefficient R) and bias. This can partly be explained by 1) the fact that additional data-assimilation was performed for the HARMONIE runs and 2) the coarser resolution of the HARMONIE model does not result in (stochastic) turbulent fluctuations in the wind field. The presented error statistics in Figure \ref{fig:scatter-validation} and the differences between the LES and HARMONIE models are representative for a larger set of validation locations, as was shown in \citet{Baas2024_wins50}. Furthermore, \citet{Baas2024_wins50} also shows that the modeled year, 2020, was particularly windy, with average wind speeds roughly 6\% higher than average. In the remainder of the paper, no corrections have made for this fact, so the presented numbers should not be considered as representative for a longer period.

\begin{figure*}[!t]
    \centering

    \begin{subfigure}{0.44\textwidth}
        \centering
        \includegraphics[
            trim={0.3cm 0 0.3cm 0},
            clip,
            width=\linewidth
        ]{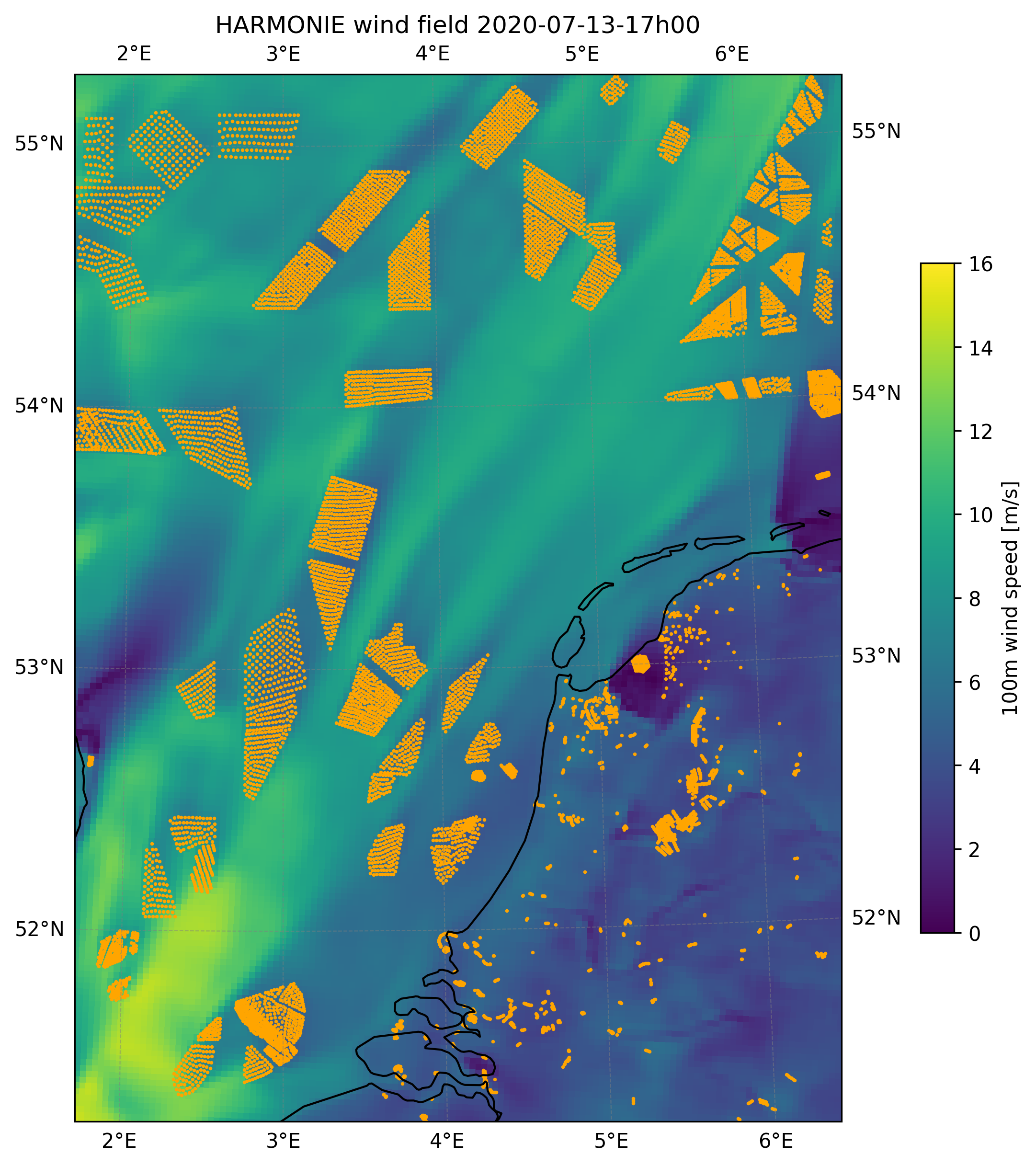}
        \caption{HARMONIE}
        \label{fig:HARMONIE-field}
    \end{subfigure}
    \hfill
    \begin{subfigure}{0.44\textwidth}
        \centering
        \includegraphics[
            trim={0.3cm 0 0.3cm 0},
            clip,
            width=\linewidth
        ]{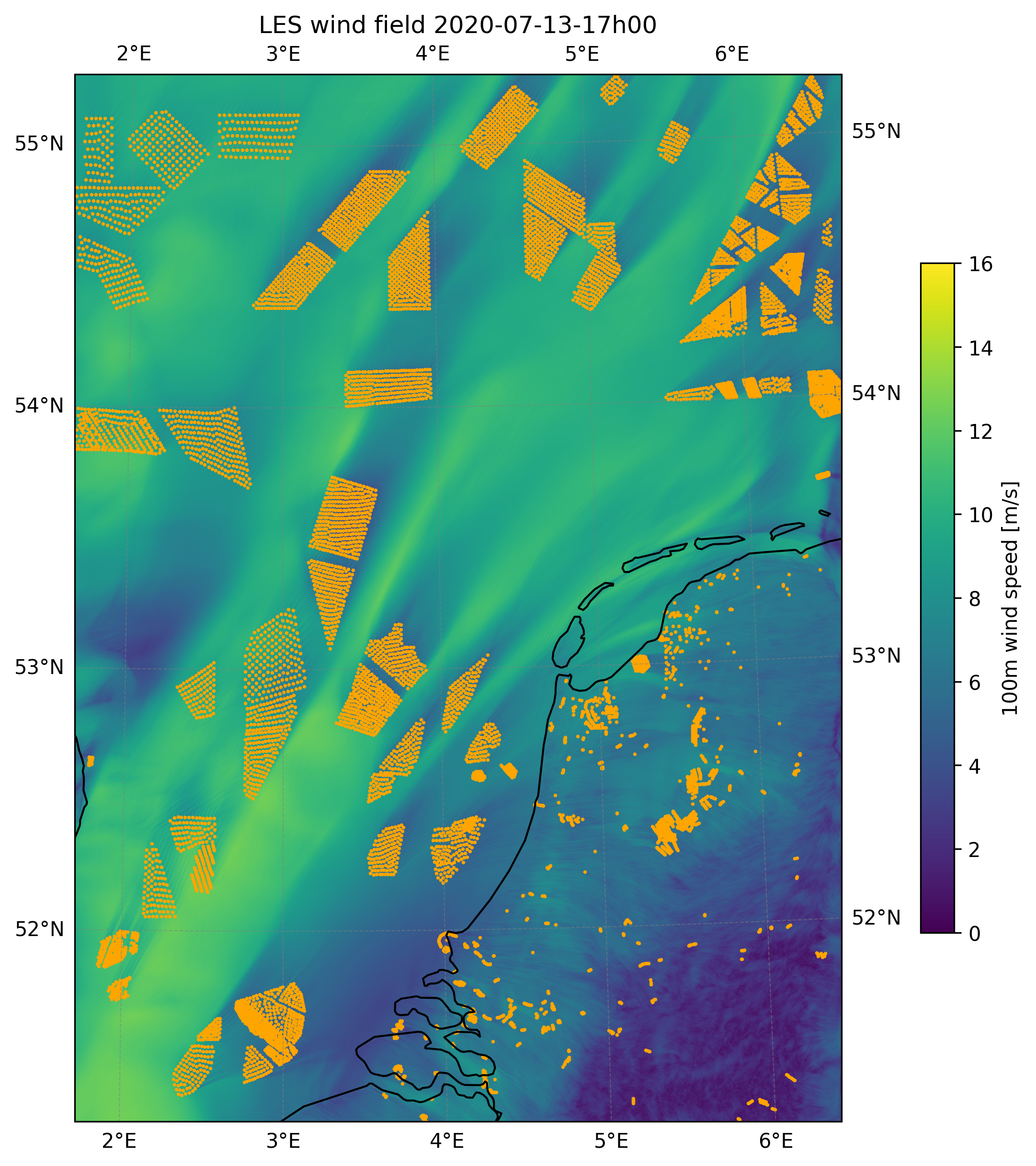}
        \caption{LES}
        \label{fig:LES-field}
    \end{subfigure}

    \par\medskip

    \begin{subfigure}{0.44\textwidth}
        \centering
        \includegraphics[
            trim={0.3cm 0 0.3cm 0},
            clip,
            width=\linewidth
        ]{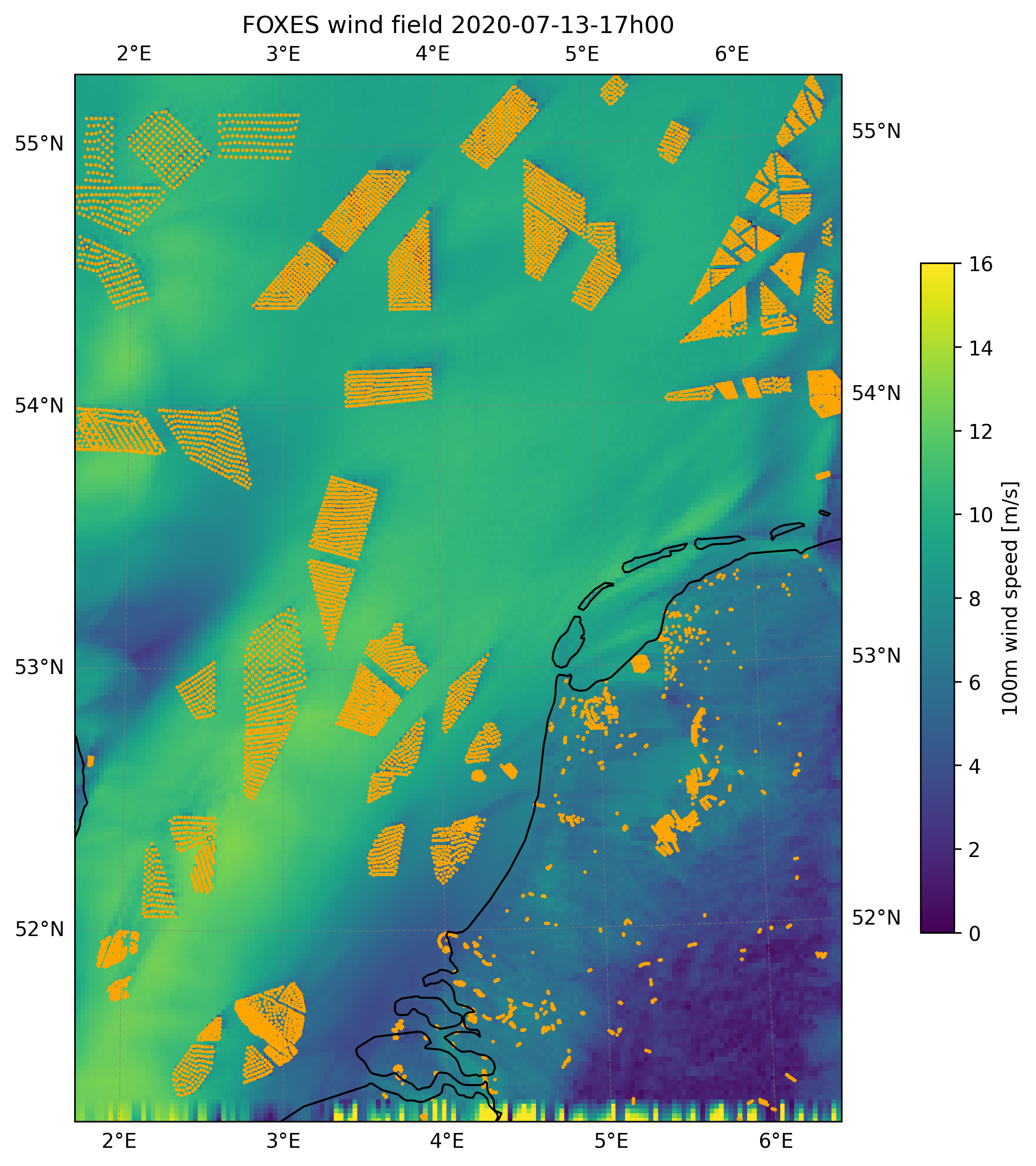}
        \caption{FOXES Model A}
        \label{fig:foxes-field}
    \end{subfigure}

    \caption{Hourly averaged wind fields of the different models
    for 2020-07-13 1700h UTC. Note that some numerical artifacts
    in the FOXES field at the lower domain boundary should be ignored.}
    \label{fig:fields}
\end{figure*}

The 2050 simulations include large amounts of hypothetical wind farms that produce significant wake effects. As a visual illustration of the typical extent and shape of the wake effects, Figure \ref{fig:fields} shows the hourly averaged 100m-wind fields for the 2050 simulation for all three models for a particular hour in summer. The offshore weather conditions on the chosen day (2020-07-13 1700h UTC) can be classified as a stable atmospheric boundary layer with southernly winds flowing over relatively cold sea water.

Figure \ref{fig:fields} reveals significant differences between the models. The wake effects visible in the HARMONIE and LES fields are comparable in shape and length, but the LES shows sharper gradients at the edges of the wakes and more pronounced speed-up effects between wind farms. The engineering model, model A (Jensen) in this case, shows much less pronounced far wake effects. In fact, at the scale of this map, it is difficult to even observe the wake effects behind the wind farms, although the internal wake effects are present.

\begin{figure*}
\centering
\includegraphics[width=0.9\linewidth]{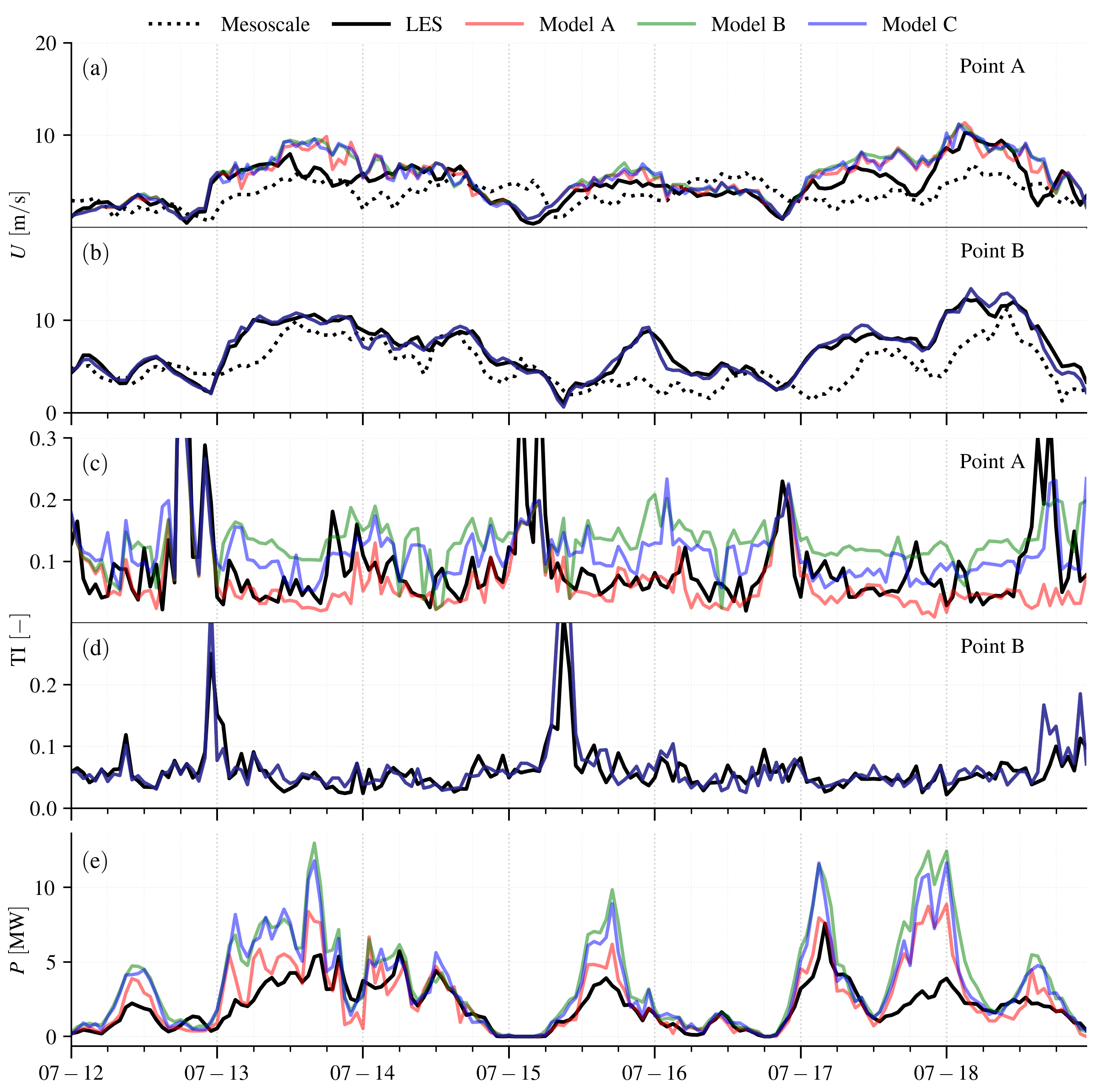}
\caption{\label{fig:time-series} Weekly time series in mid-July for the WINS50 2050 scenario comparing mesoscale, LES, and engineering wake simulations (Models A–C). Panels (a)–(b) show wind speed at 100 m at Points A and B (see Figure~\ref{fig:domain_overview}), panels (c)–(d) show turbulence intensity, and panel (e) shows the absolute power of the turbine closest to Point A. No TI or power is shown for the HARMONIE model, since those variables are not available per turbine location}
\end{figure*}

Figure~\ref{fig:time-series} provides temporal context for this event by comparing one week at two analysis locations (Figure~\ref{fig:domain_overview}). Point~A lies inside a large farm in the 2050 layout, whereas Point~B is remote from the farms added in that scenario. At Point~A, the models exhibit the same broad temporal evolution, but the engineering-model wind speeds exceed those from the LES and HARMONIE--AROME during several periods, with differences reaching approximately $5~\mathrm{ms^{-1}}$ on 13~July, including the hour shown in Figure~\ref{fig:fields}. At Point~B, Models~A--C closely follow the LES because they are driven by the 2020 LES background and the point is not strongly affected by the added farms.

The TI series at Point~A further distinguish the engineering configurations. Model~B generally predicts the highest TI, whereas Model~C also exceeds the 2050 LES during much of the week but shows smaller and less persistent differences. Model~A has no wake-added-TI formulation; its TI therefore follows the 2020 LES background supplied to the engineering model. At Point~B, the LES and engineering-model TI values remain similar except during short transient peaks. Consistent with the higher engineering-model wind speeds at Point~A, the instantaneous turbine power in Figure~\ref{fig:time-series}e is generally higher for Models~A--C during the principal production peaks, with the largest differences occurring for Model~B.  

\subsection{Annual capacity-factor patterns}

\begin{figure*}
    \centering
    \includegraphics[width=1.0\linewidth]{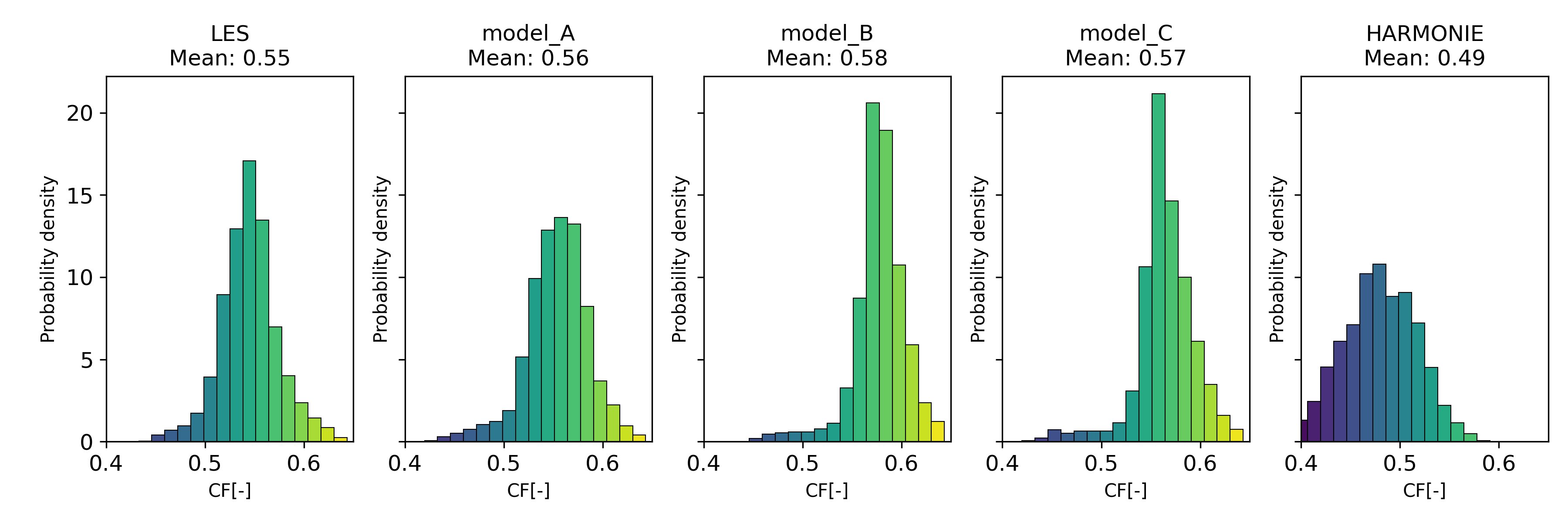}
    \caption{Histograms of the yearly capacity factors per turbine (or wind farm grid point for HARMONIE) for all models. The mean yearly capacity factors over all offshore wind farms is shown above the plots.}
    \label{fig:CF-histograms}
\end{figure*}

To get a first impression on the differences in the predicted annual capacity factors, Figure \ref{fig:CF-histograms} shows the means and probability density functions of the capacity factor for all the models in the 2050 scenario. The LES results show a mean CF of 0.55 and a fairly symmetric distribution around it. Engineering model A shows a similar distribution, while models B and C are more skewed with the majority of the data-points towards higher CF values. The HARMONIE model shows a substantially lower CF for all wind farms, with an average capacity factor of 0.49.

\begin{figure*}
    \centering
    \begin{subfigure}{0.49\textwidth}
        \includegraphics[width=\textwidth]{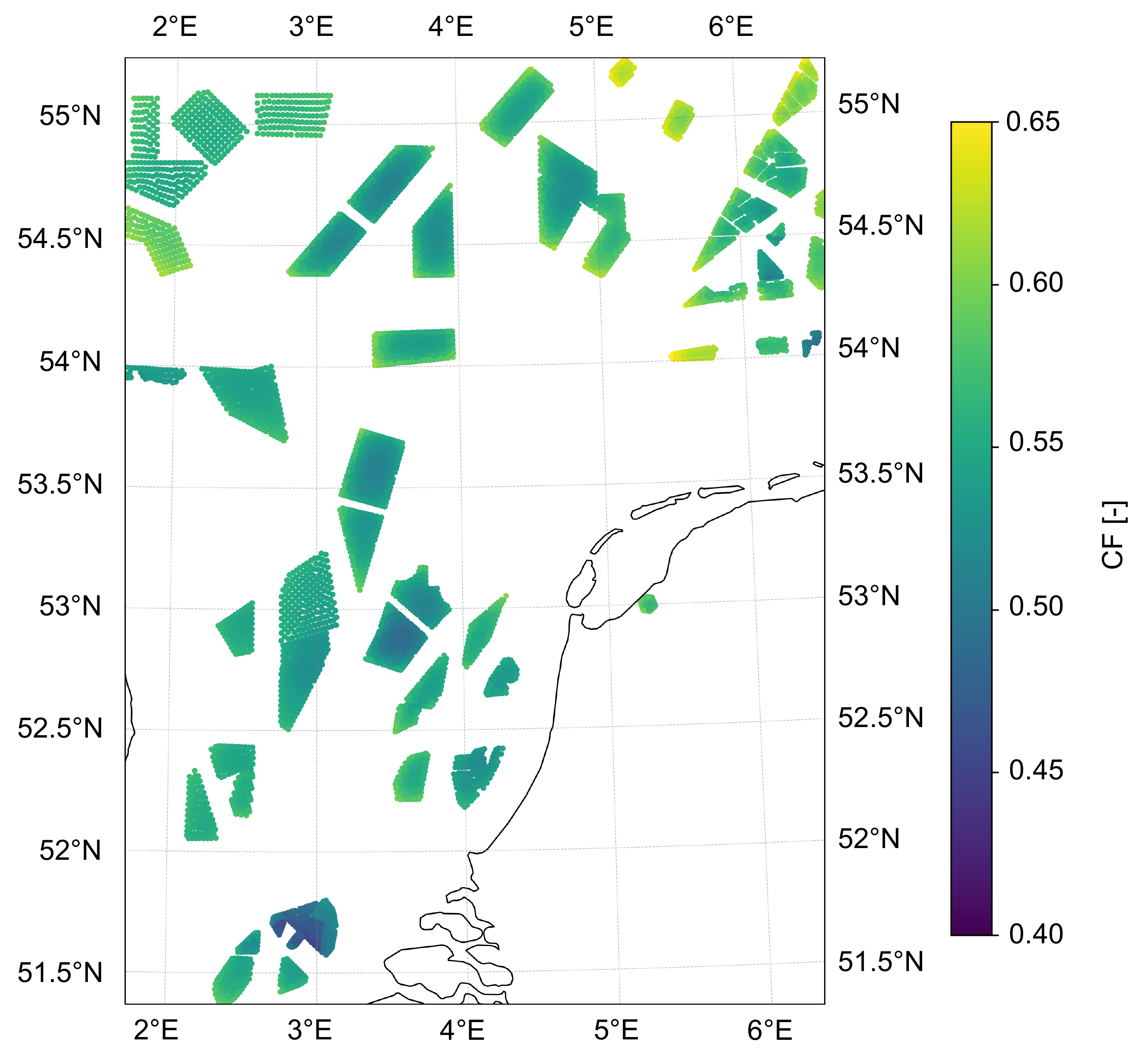}
        \caption{CF LES}
        \label{fig:CF-LES}
    \end{subfigure}
    \hfill
    \begin{subfigure}{0.49\textwidth}
        \includegraphics[width=\textwidth]{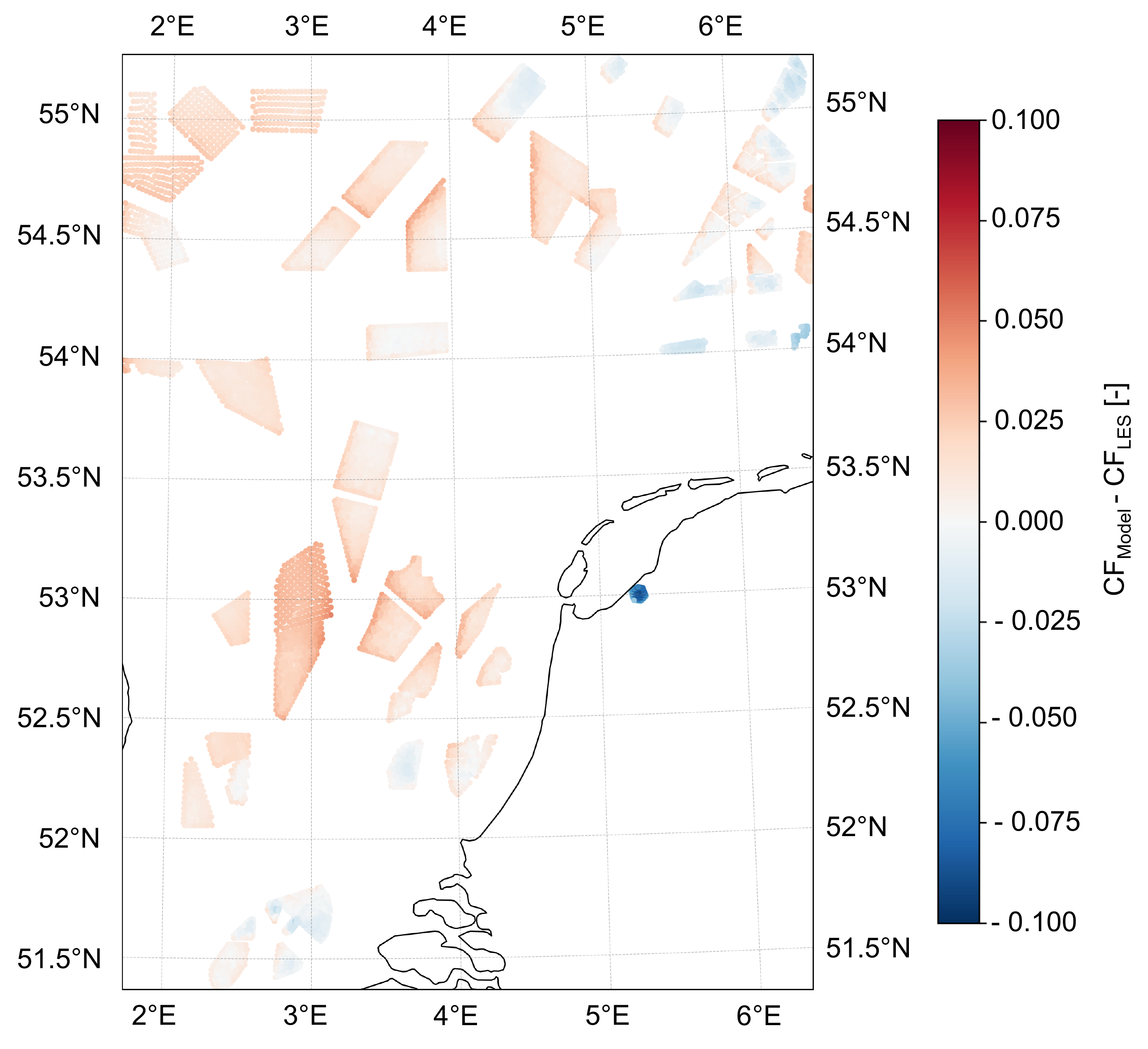}
        \caption{CF Model A - CF LES}
        \label{fig:CF-modelA}
    \end{subfigure}
    
    \begin{subfigure}{0.49\textwidth}
        \includegraphics[width=\textwidth]{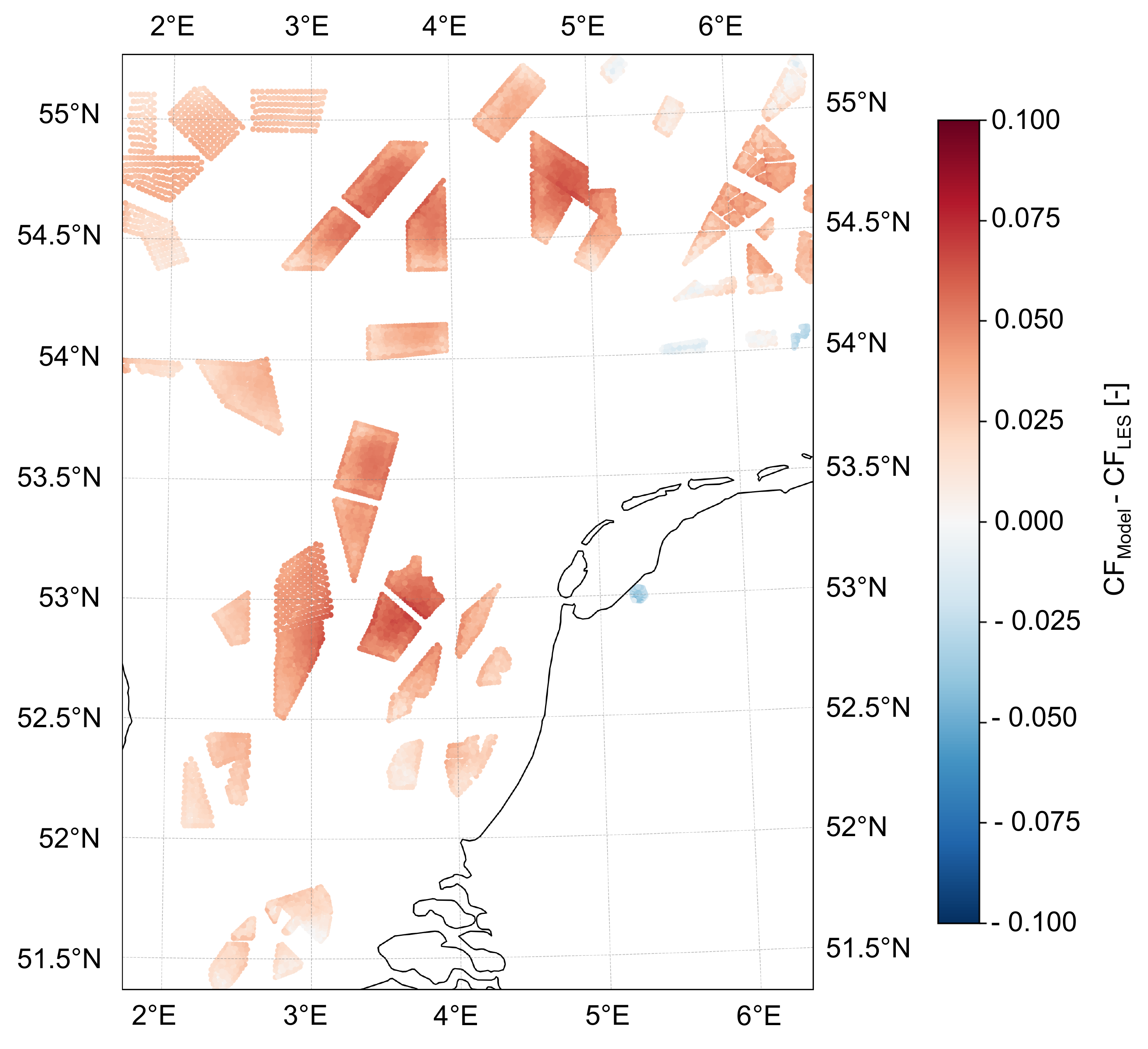}
        \caption{CF Model B - CF LES}
        \label{fig:CF-modelB}
    \end{subfigure}
    \hfill
    \begin{subfigure}{0.49\textwidth}
        \includegraphics[width=\textwidth]{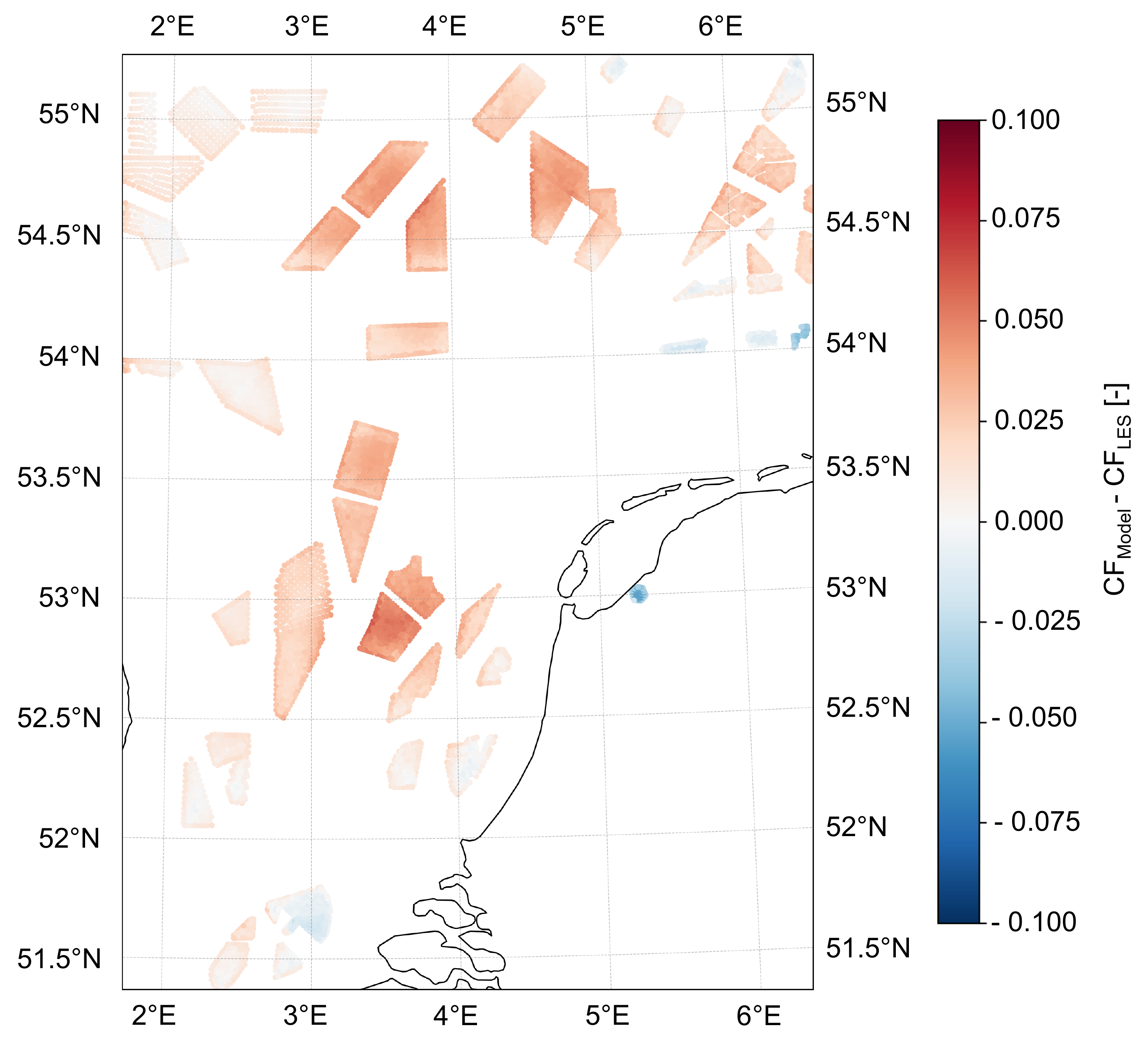}
        \caption{CF Model C - CF LES}
        \label{fig:CF-modelC}
    \end{subfigure}
    
    \caption{\label{fig:CF} Map of the capacity factor per turbine (colored dots) for the LES and the differences between LES and the three different models in the 2050 scenario}
    \label{fig:figures}

\end{figure*}

The spatial patterns of annual energy production can be assessed by looking at the annual capacity factors ($CF=P_{avg}/P_{nom}$) per individual turbine. Figure \ref{fig:CF} shows maps of the CF all individual turbines for the LES (a) and the differences between the three engineering models with respect to the LES CF (b to d) in the 2050 scenario. The CF computed in the LES runs (Figure \ref{fig:CF-LES}) shows that values range roughly between 0.4 and 0.65, the highest values being found in smaller isolated wind farms in or near the German Bight, where wind speeds are generally the highest. The lowest values are found in the middle of the large additional wind farm zones and in the Borssele offshore cluster (in the Southwest part of the domain at 51.7$^\circ$N, 3$^\circ$E). 

The differences in CF between LES and the three engineering models show pronounced spatial patterns. Overall, Model~A (Jensen, no added TI) shows the smallest differences. However, Figure~\ref{fig:CF-modelA} shows a clear overestimation of CF at the edges of large additional wind farms. This overestimation is likely related to the absence of global blockage and the underestimation of external wakes from neighboring wind farms. Model~B (Linear Gaussian, Crespo--Hernandez TI) predicts substantially higher CF values than LES, especially in the centers of large wind farms (Figure~\ref{fig:CF-modelB}). This indicates an underestimation of wake effects for the particular model settings used. Model~C (TurbOPark Gaussian, Frandsen TI) shows a more nuanced picture, with an overestimation of CF for large wind farms but values closer to LES results for smaller wind farms. Furthermore, Figure~\ref{fig:CF-modelC} shows mixed results regarding where wake effects are strongest. For some farms, such as IJmuiden Ver, differences with respect to LES are highest in the wind farm center, whereas for others, such as Hollandse Kust West, they are largest at the edges.

\begin{figure*}
    \centering
    \begin{subfigure}{0.49\textwidth}
        \includegraphics[width=\textwidth]{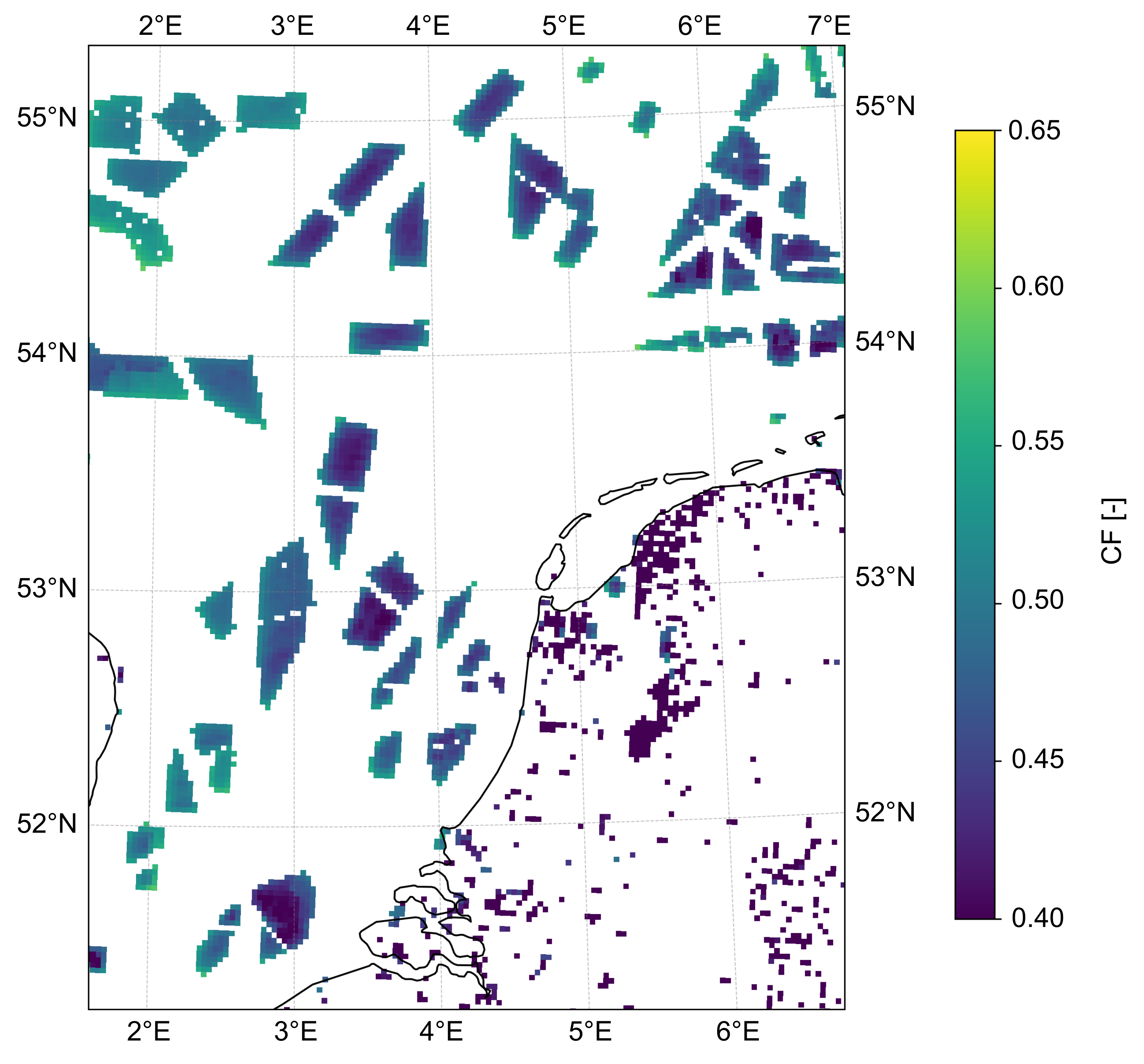}
        \caption{CF HARMONIE}
        \label{fig:cf-harmonie}
    \end{subfigure}
    \hfill
    \begin{subfigure}{0.49\textwidth}
        \includegraphics[width=\textwidth]{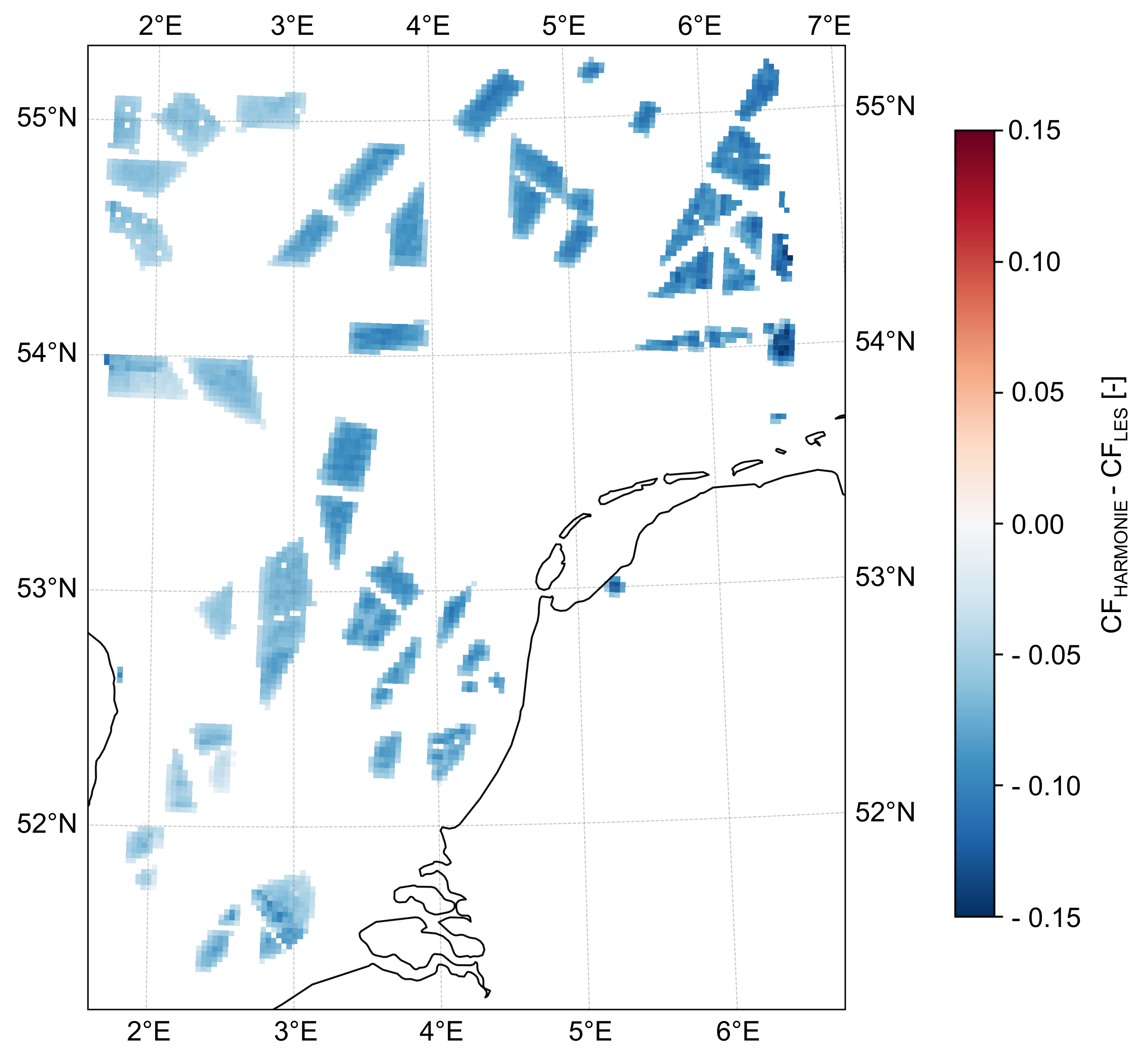}
        \caption{CF HARMONIE - CF LES}
        \label{fig:delta-cf}
    \end{subfigure}
    \caption{Capacity factor of the HARMONIE model (left) and difference in capacity factor between HARMONIE and LES (right). Note that LES CF values per turbine have been interpolated to the HARMONIE grid and that there are small differences in the turbine locations in HARMONIE and LES.}
    \label{fig:cf-harmonie-les}
\end{figure*}

Figure \ref{fig:cf-harmonie-les} shows the capacity factors as modeled by the HARMONIE mesoscale model and the differences with respect to the LES results. Because the mesoscale model does not resolve individual turbines but calculates power per grid cell, the LES results have been interpolated to the HARMONIE grid, but only for offshore turbines since there were small differences in the modeled onshore turbines.

The HARMONIE model shows lower capacity factors than the LES for all wind farms. The differences are likely the result of several factors and we emphasize that no conclusions can be drawn from this paper which model is more accurate in terms of modeled capacity factors. First, as shown in Figure \ref{fig:scatter-validation} the LES calculates higher free-stream wind speeds, which translates automatically into higher capacity factors, even in the absence of wake effects. Secondly, the Fitch parameterization used in the HARMONIE runs did not take the later modification proposed by \citet{Vollmer2024} into account, thus possibly overestimating wake effects. Third, \citet{Baas2024_wins50} showed that in terms of wake effects (or, equivalently, velocity deficits), the CF differences are much smaller since they are not sensitive to the differences in free-stream wind speed. 

Finally, the CF values presented in Figures~\ref{fig:CF} and \ref{fig:cf-harmonie-les} are based directly on the model outputs; neither a long-term adjustment nor a bias correction was applied. Because 2020 was a particularly windy year, with wind speeds approximately 6\% above the long-term average \citep{Baas2024_wins50}, the absolute CF values may exceed climatological expectations. Such corrections are beyond the scope of this study, which focuses on comparing model estimates under a common meteorological year.

\subsection{Dependence of model errors on distance from wind farms}

\begin{figure*}
    \centering
    \includegraphics[width=0.80\linewidth]{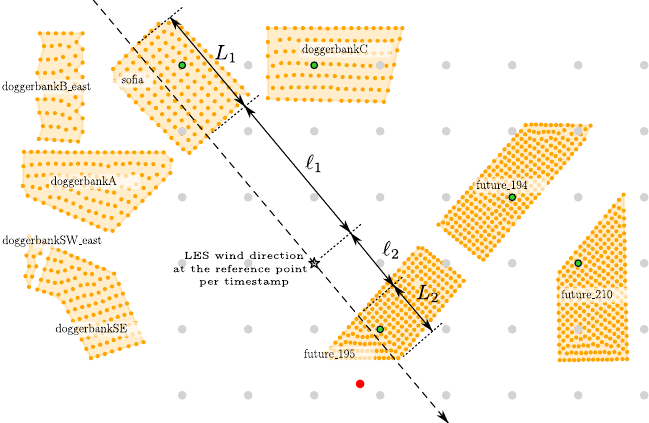}
    \caption{Definition of the wind-farm-relative distances for an example north-westerly flow at the reference point (star). The upwind ray defines the boundary distance $l_1$ and polygon chord $L_1$; the downwind ray defines $l_2$ and $L_2$. A point enters the inter-farm blockage class when the downwind polygon represents a farm added in the 2050 scenario, $l_2/L_2<0.2$, and either no upwind polygon is intercepted or $l_1/L_1>20$. The class therefore identifies points outside wind farmpolygons that are immediately upstream of an added farm and have no nearby intercepted upwind farm. Because the first alternative permits no upwind interception, ``inter-farm'' does not require every classified point to lie geometrically between two farms. The classification is geometric and is not an isolated measure of blockage.} 
    \label{fig:distances}
\end{figure*}

For each hour of the year, the evaluation points are classified using the wind direction from the 2020 LES background. Straight rays cast from each point in the upwind and downwind directions identify the first intercepted wind farm polygons (Figure~\ref{fig:distances}). The distance from the upwind-farm boundary to the point is denoted by $l_1$, and the corresponding polygon chord by $L_1$. Similarly, $l_2$ is the distance from the point to the downwind-farm boundary and $L_2$ is the associated chord. The plotted classes distinguish an inter-farm blockage region, points within farms added in the 2050 scenario and successive inter-farm wake-distance intervals. Here, "inter-farm" denotes points outside wind farm polygons; it does not imply that every point lies between intercepted farms on both sides.

The downstream-distance classes include only samples for which the nearest intercepted upwind polygon represents a farm added in the 2050 scenario. This prevents distances to farms operating in 2020, whose wakes are already embedded in the common background fields, from being mixed into the conditional statistics. Evaluation points inside those farms are likewise excluded from the plotted classes. Biases are pooled directly over the valid point-hour samples in each class, without prior averaging by point or hour. Only Models~A--C are compared with the LES because they use the same background fields and turbine definitions. A HARMONIE--AROME comparison would additionally contain differences in atmospheric forcing, wind direction and spatial representation.

\begin{figure*}
    \centering
    \includegraphics[width=1\linewidth]{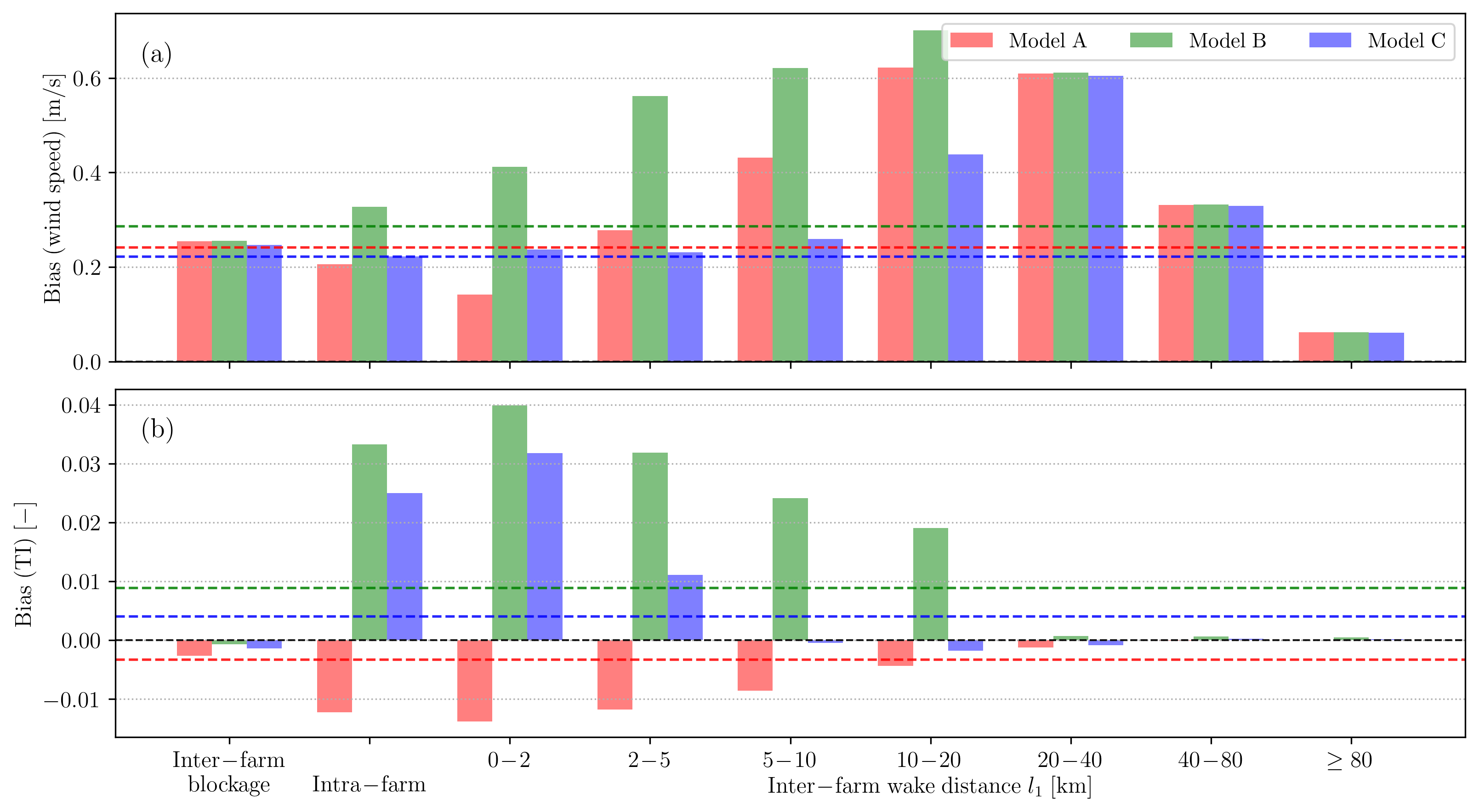}
    \caption{Mean engineering-model bias relative to the LES in (a) wind speed at 100~m and (b) turbulence intensity (TI), conditioned on wind-farm-relative position. The first class is defined in Figure~\ref{fig:distances}; the intra-farm class contains points inside farms added in the 2050 scenario. The downstream classes contain only samples whose nearest intercepted upwind polygon represents an added farm and are binned by the boundary distance $l_1$. Each bar shows the model-minus-LES mean over all valid point-hour samples in its class. The plotted classes contain 68.6\% of all valid point-hours; their shares from left to right are 0.7\%, 13.7\%, 2.0\%, 2.9\%, 3.9\%, 7.1\%, 10.9\%, 13.1\% and 14.4\% of the full valid sample. Dashed lines show the corresponding annual mean biases over all 3,276,059 valid point-hours, including samples outside the displayed classes, and are therefore not averages of the bars. Positive wind speed bias indicates a higher wind speed in the engineering model. TI is dimensionless.}
    \label{fig:bias-bar}
\end{figure*}

In the inter-farm blockage class, all three engineering models have a positive mean wind speed bias of 0.247--0.256 $\mathrm{ms^{-1}}$ (Figure~\ref{fig:bias-bar}a). Their TI biases are small and negative, ranging from -0.0027 to -0.0007. The similar wind speed biases are consistent with an upstream wind speed reduction represented by the LES but absent from the engineering configurations. They should not, however, be interpreted as an isolated estimate of blockage because the class is defined geometrically and each model--LES difference remains a composite of all differences between the corresponding simulations.

Within the added farms, the mean wind speed biases are approximately 0.21, 0.33 and 0.22 $\mathrm{ms^{-1}}$ for Models~A, B and C, respectively. Immediately downstream, at 0--2~km, the corresponding values are 0.14, 0.41 and 0.24 $\mathrm{ms^{-1}}$. The distance dependence is therefore non-monotonic and model specific. Models~A and B reach their largest biases in the 10--20~km interval, at approximately 0.62 and 0.70 $\mathrm{ms^{-1}}$, whereas Model~C reaches its maximum of approximately 0.60$\mathrm{ms^{-1}}$ at 20--40~km. Model~B has the largest bias from the intra-farm class through 10--20~km, but the three means converge to 0.60--0.61 $\mathrm{ms^{-1}}$ at 20--40~km and approximately 0.33 $\mathrm{ms^{-1}}$ at 40--80~km. At $l_1\geq80$~km, a small positive bias of approximately 0.06$\mathrm{ms^{-1}}$ remains for every model. Because the engineering models are driven by the 2020 LES background, the positive downstream biases are consistent with a weaker aggregate wind speed reduction than in the 2050 LES, but they cannot be attributed uniquely to wake recovery. The largest discrepancies occur at intermediate downstream distances rather than immediately behind the farm boundary.

The far-wake comparison must also account for the executed wake-length settings. Models~A and C used a 20~km maximum streamline length, whereas Model~B imposed no explicit maximum (see Appendix \ref{app}). Despite this difference, the conditional means of the three models are nearly identical in the 20--40~km distance class and beyond. This agreement cannot be interpreted as evidence that the models have similar intrinsic far-wake recovery because the distances are defined differently. Specifically, $l_1$ measures the straight-line distance from a wind-farm polygon boundary, whereas the 20~km limit is measured from each turbine along its two-dimensional wake streamline. Consequently, the $l_1$ classes do not indicate whether a given wake has reached the imposed streamline limit, and Figure~\ref{fig:bias-bar} cannot separate the effect of wake truncation from differences in far-wake recovery.

The TI biases provide a complementary distinction (Figure~\ref{fig:bias-bar}b). Because Model~A has no wake-added-TI formulation, its bias compares the 2020 LES background TI supplied to the model with the TI represented by the 2050 LES. Its mean bias is approximately -0.012 within farms and -0.014 at 0--2~km, and approaches zero beyond 20~km. Model~B exceeds the LES by approximately 0.033 within farms and 0.040 at 0--2~km; its positive bias decreases with distance but remains approximately 0.019 at 10--20~km before approaching zero at 20--40~km. Model~C has positive biases of approximately 0.025 within farms and 0.032 at 0--2~km, but the bias decreases to 0.011 at 2--5~km and is close to zero from 5--10~km onwards. The simultaneous positive TI and wind speed biases of Model~B within 20~km are compatible with enhanced wake recovery in that configuration, but do not establish causality. At 20--40~km, all three wind speed biases remain approximately 0.60--0.61 $\mathrm{ms^{-1}}$, whereas their TI biases are within approximately 0.0013 of zero. Wake-added TI alone therefore cannot explain the wind speed-bias profile; velocity-deficit and superposition formulations, spatial representation and wake-length treatment remain confounded.

\subsection{Monthly variations}

\begin{figure*}
    \centering
    \includegraphics[width=1.00\linewidth]{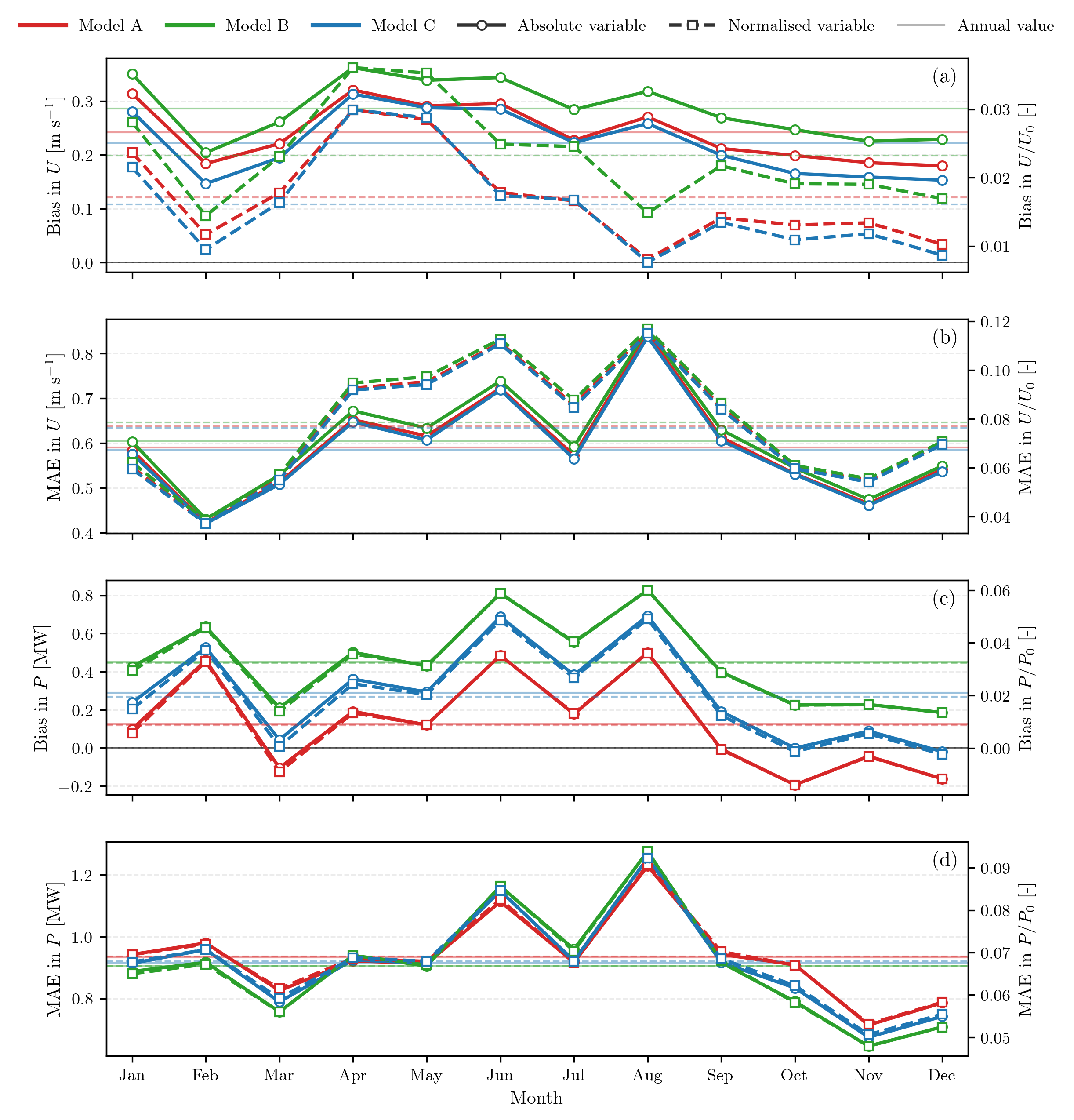}
    \caption{Monthly errors of Models~A--C relative to the LES. Panels show bias in (a) wind speed, $U$, and (c) turbine power, $P$, and the corresponding mean absolute errors (MAEs) in (b) and (d). Solid lines with circles denote dimensional quantities $U$ and $P$ (left axes), whereas dashed lines with squares denote normalised quantities $U/U_0$ and $P/P_0$ (right axes), where $P_0$ is rated power. Thin horizontal lines indicate annual values. Monthly wind speed statistics include all valid hourly samples at 373 evaluation points; power statistics include all valid hourly samples across the 6,593 turbines added in the 2050 scenario. Bias is defined as model minus LES, so positive values indicate overprediction.
}
    \label{fig:monthly_errors}
\end{figure*}

Figure~\ref{fig:monthly_errors} shows the monthly variation in domain-aggregated model--LES errors. Wind speed bias is positive for all three engineering models in every month. Thus, when averaged over all valid point-hour samples at the 373 evaluation points, the engineering models predict higher wind speeds than the LES; this does not imply a positive error at every point and hour. Model~B has the largest monthly bias, whereas Models~A and C are generally closer to one another. Across the three models and twelve months, the dimensional wind speed MAE ranges from approximately 0.42 $\mathrm{ms^{-1}}$ in February to 0.84 $\mathrm{ms^{-1}}$ in August. The dimensional and normalised wind speed MAEs exhibit similar monthly variation, with minima in February and maxima in August. The variation in dimensional MAE is therefore not caused solely by monthly differences in the magnitude of the background wind speed, $U_0$.

The configurations differ more clearly in the sign and magnitude of their power biases. Model~B has a positive bias in every month, with its largest values in June and August. Model~C is also positively biased during most of the year, but its bias approaches zero in October and becomes slightly negative in December. Model~A changes sign several times: its bias is negative in March, near zero to slightly negative in September, and negative from October to December. The small positive annual bias of Model~A therefore masks opposing monthly contributions. Model~C undergoes less cancellation among monthly mean biases because it remains positively biased through most of the year.

In months for which mean power bias approaches zero, power MAE remains approximately $0.8$--$0.9~\mathrm{MW}$. Positive and negative turbine-hour errors must therefore partly cancel in the monthly mean, and the annual bias may contain additional cancellation among months. Bias and MAE consequently provide complementary information: bias describes the signed mean error, whereas MAE describes its mean magnitude. Overall, Model~B has the largest positive annual bias and remains positively biased throughout the year. Model~A has a smaller positive annual bias but exhibits several monthly sign reversals, whereas Model~C has an intermediate annual bias and remains positive in most months.

The sign changes in power bias, particularly for Models~A and C, show that aggregation over the year can conceal opposing monthly contributions. Because the occurrence of wind speed and stability regimes also varies seasonally, we next examine whether the farm-level power differences are systematically associated with atmospheric conditions. This conditional analysis does not, however, determine whether seasonal changes in those conditions explain the monthly variation.

\subsection{Model differences in relation to atmospheric conditions}

For each of the 31 farms with a representative virtual mast, turbine power is first summed and divided by the farm's installed capacity to obtain an hourly capacity factor. Each farm-hour is paired with the 100~m wind speed, $U$, and bulk Richardson number, $Ri$, from the 2020 LES background at the corresponding mast. The bulk Richardson number is calculated from differences in virtual potential temperature and the horizontal wind vector between 50 and 150~m. A common finite-data mask is applied to the LES and all three engineering-model configurations. Samples are retained when $4\leq U<25~\mathrm{ms^{-1}}$ and $|Ri|\leq2$, and the figure displays four 2~$\mathrm{ms^{-1}}$ intervals between 6 and $14~\mathrm{ms^{-1}}$. Stability classes are defined as unstable ($Ri<-0.05$), near-neutral ($-0.05\leq Ri<0.05$), stable ($0.05\leq Ri<0.25$) and strongly stable ($Ri\geq0.25$).

\begin{figure*}[h!]
    \centering
    \includegraphics[width=1.00\linewidth]{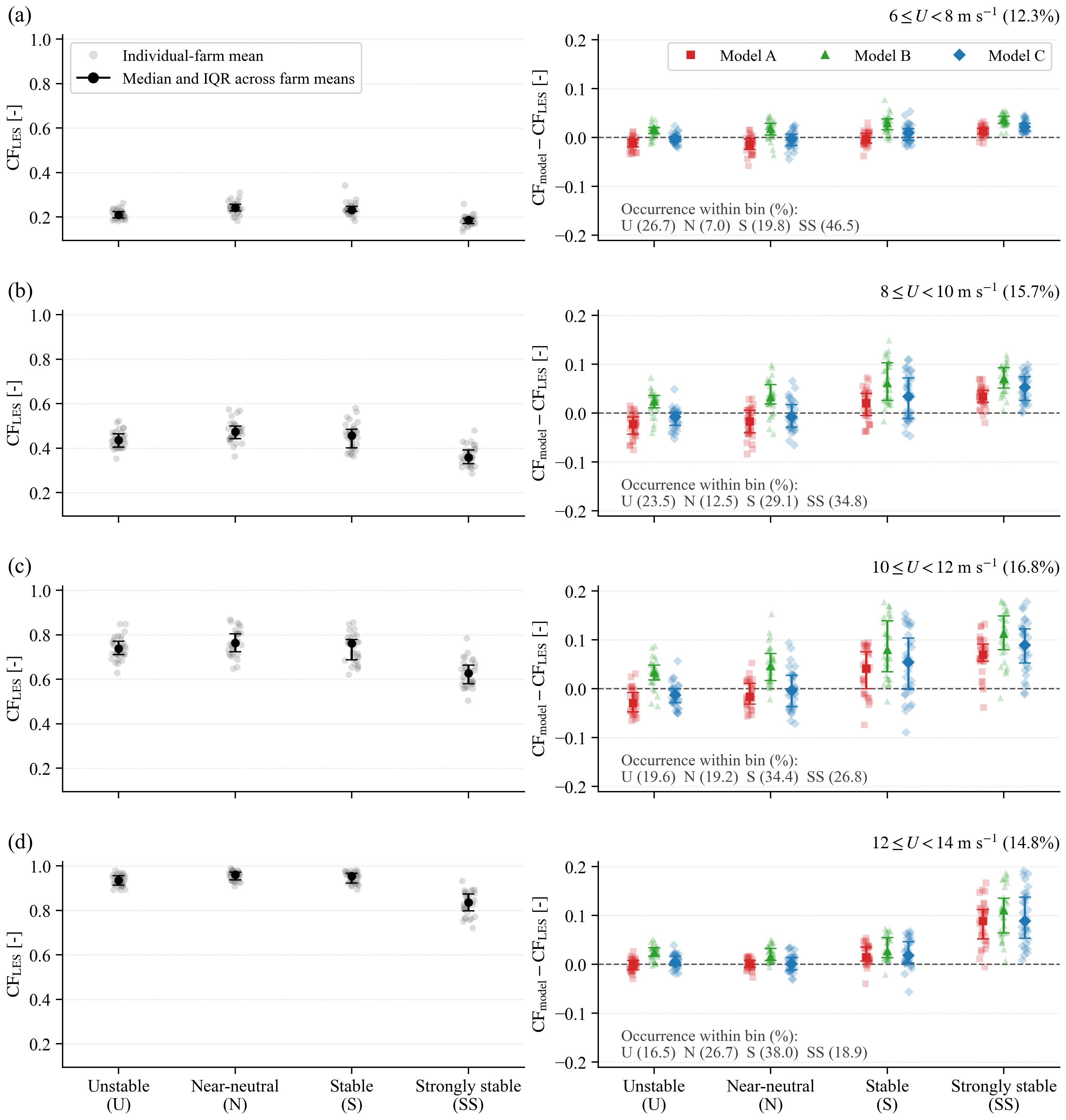}
    \caption{Farm-level capacity factor and engineering-model bias conditioned on 100~m wind speed and bulk Richardson number from the 2020 LES background. Rows (a)--(d) correspond to $6\leq U<8$, $8\leq U<10$, $10\leq U<12$ and $12\leq U<14~\mathrm{ms^{-1}}$, respectively. The percentage beside each row heading is the occurrence of that wind speed interval among the 199,353 common valid farm-hours satisfying $4\leq U<25~\mathrm{ms^{-1}}$ and $|Ri|\leq2$. The percentages within each right-hand panel give the conditional occurrence of the four stability classes within that wind speed interval. Left panels show LES capacity factor; right panels show the model-minus-LES capacity-factor bias. Light symbols are arithmetic means for individual farms. Dark symbols and error bars are the median and IQR across the 31 farm means. Each farm--wind speed--stability cell contains at least 10 farm-hours. Bias and capacity factor are dimensionless.}
    \label{fig:stability}
\end{figure*}

To give each farm equal weight, an arithmetic mean is calculated separately for each farm within every wind speed--stability cell. Figure~\ref{fig:stability} shows those individual-farm means together with their median and interquartile range (IQR) across farms. The resulting summaries are therefore farm-balanced rather than weighted by installed capacity or sample count. A minimum of 10 farm-hours is required for a farm to enter a cell. All 31 farms satisfy this requirement in all 16 displayed cells; even the least populated farm--cell contains 25 hours. The four displayed wind speed intervals contain 118,809 farm-hours, corresponding to 59.6\% of the 199,353 farm-hours that satisfy the general filters.

LES capacity factor increases with wind speed in all stability classes, but strongly stable conditions are associated with consistently lower values within each wind speed interval (Figure~\ref{fig:stability}, left column). The cross-farm median LES capacity factors under near-neutral and strongly stable conditions are 0.242 and 0.186 at 6--8~$\mathrm{ms^{-1}}$, 0.475 and 0.359 at 8--10~$\mathrm{ms^{-1}}$, 0.763 and 0.628 at 10--12~$\mathrm{ms^{-1}}$, and 0.961 and 0.835 at 12--14~$\mathrm{ms^{-1}}$. The differences between these plotted medians are therefore 0.056, 0.115, 0.135 and 0.125, respectively. In contrast, the unstable, near-neutral and stable medians are comparatively close within each interval. The conditional stability distribution also changes with wind speed: strongly stable conditions account for 46.5\% of the 6--8~$\mathrm{ms^{-1}}$ bin but 18.9\% of the 12--14~$\mathrm{ms^{-1}}$ bin, whereas the stable fraction increases from 19.8\% to 38.0\%.

At 6--8~$\mathrm{ms^{-1}}$, the median model--LES differences are small, ranging from approximately $-0.014$ to 0.036 across the models and stability classes. The dependence on atmospheric regime becomes more pronounced at 8--12~$\mathrm{ms^{-1}}$. Model~B has a positive cross-farm median bias in every displayed wind speed--stability cell; these medians range from 0.015 to 0.113, and the lower quartile remains above zero in all 16 cells. Models~A and C have negative median biases in every unstable and near-neutral cell between 6 and 12~$\mathrm{ms^{-1}}$. Their medians become positive under stable and strongly stable conditions at 8--12~$\mathrm{ms^{-1}}$; at 6--8~$\mathrm{ms^{-1}}$, the stable median remains slightly negative for Model~A. At 10--12~$\mathrm{ms^{-1}}$, for example, the strongly stable median biases are 0.070, 0.113 and 0.090 for Models~A, B and C, respectively.

At 12--14~$\mathrm{ms^{-1}}$, the LES median capacity factor is close to one under unstable, near-neutral and stable conditions. The corresponding median biases of Models~A and C are close to zero or modestly positive, whereas Model~B retains positive medians of 0.024, 0.015 and 0.027 in these three classes. Under strongly stable conditions, however, the LES median capacity factor decreases to 0.835, and all three engineering-model configurations have appreciable positive median biases: 0.089, 0.110 and 0.089 for Models~A, B and C, respectively.

The conditioning therefore reveals two recurring features in the cross-farm medians that are obscured by annual aggregation. Model~B has a positive median bias in every displayed atmospheric regime, whereas Models~A and C change from negative median biases under unstable and near-neutral conditions to positive biases under stable and strongly stable conditions, most clearly at 8--12~$\mathrm{ms^{-1}}$. These regime-dependent sign changes provide a possible mechanism for cancellation in aggregated biases. However, the stability analysis uses equal-farm summaries of capacity-factor differences, whereas the monthly statistics pool turbine-hour errors; it therefore does not provide a quantitative decomposition of the monthly or annual biases. The observed associations should not be interpreted as the isolated causal effect of atmospheric stability because wind direction, within-bin wind speed variation, farm geometry, background turbulence and the nonlinear power curve remain coupled to the stability classes. Nor does the annual conditional analysis establish that seasonal changes in stability occurrence cause the monthly variations in Figure~\ref{fig:monthly_errors}.

\section{Conclusions and future work} 
\label{sec:conclusions}

This study presents a regional comparison over a full year of engineering wake models, mesoscale atmospheric modeling, and large-eddy simulation (LES) for a 100.7~GW offshore wind deployment scenario in the North Sea for 2050, using meteorological conditions from 2020. These approaches represent different levels of atmospheric and wake dynamics. HARMONIE--AROME and ASPIRE dynamically represent the evolution of the atmosphere while accounting for turbine momentum extraction, whereas the engineering models describe turbine wakes through computationally efficient reduced order formulations applied to a prescribed background flow. Both HARMONIE--AROME and the ASPIRE LES reproduced the observed offshore wind variability, with smaller pointwise errors for HARMONIE--AROME at the validation sites. This agreement provides confidence in the atmospheric conditions used for the subsequent analysis. The comparison then combined a descriptive assessment across the three modeling approaches with a more controlled evaluation of three engineering configurations using common background fields derived from LES and consistent turbine definitions.

The results show that both regional wake characteristics and estimates of annual production depend on the modeling approach. Instantaneous flow fields and time series indicate substantially stronger far wake effects in HARMONIE--AROME and LES than in the engineering configurations. These differences are also reflected in annual production, with capacity factors ranging from approximately 0.49 in HARMONIE--AROME to 0.58 in the Gaussian engineering configuration with the lowest estimated losses. Within the controlled comparison at turbine level, Model~A produced an annual capacity factor distribution similar to that of LES, whereas Models~B and C generally predicted higher values. The spatial structure of these differences, however, was not uniform. Model~A showed its largest deviations from LES near the edges of large wind farms, while the differences for Models~B and C were more pronounced toward the farm interiors. However, the lower grid level capacity factors from HARMONIE--AROME are not directly equivalent to these turbine level estimates and should not be interpreted as evidence of systematically better or worse model performance. More generally, agreement with LES in annual capacity factor does not necessarily imply agreement in the spatial structure or physical evolution of the wakes.

The controlled comparison between the engineering models and LES makes this distinction particularly clear. All three engineering configurations predicted higher domain averaged wind speeds than LES in every month. The largest differences occurred approximately 10--40~km downstream of the added farms rather than directly behind the farm boundaries. Differences were also present immediately upstream, where values of approximately 0.25~m~s$^{-1}$ were consistent with the upstream flow reduction represented by LES but absent from the engineering configurations. This signal provides evidence of a regional blockage response. However, the upstream comparison alone should not be interpreted as an isolated measure of global blockage. Differences within the wind farms were of a similar order. Farther downstream, however, the departure became more pronounced as the engineering models recovered toward the background flow more rapidly than LES. This behavior is consistent with the stronger far wake signatures seen in both LES and HARMONIE--AROME and indicates that the largest discrepancies are not confined to turbine scale or internal farm wakes. Differences in wake added turbulence contributed to this behavior but did not explain the downstream wind speed differences on their own. Taken together, these spatial patterns show that regional wake model evaluation must extend beyond individual wind farms and their immediate wakes to include flow recovery over tens of kilometers.

The corresponding power differences also varied systematically with time and atmospheric regime. Model~B remained positively biased across months and across the wind speed and stability classes. Models~A and C, in contrast, exhibited monthly changes in the sign of their biases and shifted from negative biases under unstable and near neutral conditions to positive biases under stable and strongly stable conditions. The largest discrepancies occurred when high wind speeds coincided with stable atmospheric conditions. Strongly stable conditions were also associated with the lowest LES capacity factors and the clearest common overprediction by the engineering configurations. These patterns indicate that differences between the models cannot be described by a single annual correction or bias. Moreover, small aggregated power biases often coincided with substantial absolute errors because positive and negative turbine hour differences compensated when averaged over a month or year. Agreement in annual energy production can therefore conceal important differences in when, where, and under which atmospheric conditions power losses occur.

These findings have direct implications for the use of engineering wake models in dense regional deployments. Model choice affects not only estimated annual production but also the predicted spatial extent of wake effects and their sensitivity to atmospheric conditions. Sensitivity to model formulation should therefore be reported alongside regional energy yield estimates, while model validity should be assessed across downstream distance and atmospheric regime rather than from aggregated production alone. At the same time, engineering wake models remain essential for efficiently exploring the large number of layouts and operating conditions required in regional wind energy assessments. The present comparison identifies global blockage, far wake recovery, and stable atmospheric conditions as areas where their simplified representation of atmospheric feedbacks can lead to systematic differences from LES. Improving the representation of these processes while retaining computational efficiency remains an important direction for engineering wake modeling. Data driven wake formulations and emerging machine learning approaches to atmospheric modeling may provide complementary pathways toward this objective. Their value, however, will depend on whether they can reproduce the relevant offshore wake behavior across the range of atmospheric conditions that control regional wake losses.

Several limitations define the scope of these conclusions. The analysis characterizes model dependence for a single, unusually windy meteorological year and should not be interpreted as a climatological estimate of 2050 energy production. LES is also used as a more process resolving numerical reference for the controlled comparison with the engineering models rather than as observational ground truth. In addition, differences among the three broader modeling approaches reflect not only wake formulation but also differences in numerical representation and model execution. Future work should use increasingly harmonized model settings to separate formulation effects from implementation choices, extend the comparison across multiple meteorological years, and evaluate wake predictions at the scale of wind farm clusters against observations. Moreover, it should assess systematically the sensitivity of each engineering-model configuration to its empirical parameters and component choices, including wake-recovery rates and superposition methods. Targeted simulations in which individual farms or clusters are selectively removed could further distinguish internal wake losses, external wakes, and regional blockage effects. In general, the results show that similar annual energy estimates can arise from substantially different representations of the underlying flow. Establishing where these differences occur and under which atmospheric conditions they become important is essential for assessing the realizable energy production of future large scale offshore wind development.

\section*{CRediT authorship contribution statement}

\textbf{Jhon J. Quiñones:} Methodology, Software, Formal analysis, Investigation, Visualization, Data curation, Writing -- original draft, Writing -- review \& editing.
\textbf{Diego Siguenza-Alvarado:} Methodology, Software, Formal analysis, Visualization, Data curation, Writing -- original draft, Writing -- review \& editing.
\textbf{Peter Baas:} Software, Conceptualization, Investigation, Data curation, Resources,
Visualization, Writing -- review \& editing.
\textbf{Pim van Dorp:} Software, Investigation, Data curation, Resources, Visualization, Writing -- review \& editing.
\textbf{Majid Bastankhah:} Methodology, Supervision, Writing -- review \& editing.
\textbf{Remco A. Verzijlbergh:} Conceptualization, Methodology, Software, Investigation, Data curation, Resources, Visualization, Supervision, Writing -- review \& editing.

\section*{Declaration of generative AI and AI-assisted technologies in the writing process}

During the preparation of this work the authors used Chat GPT-5o and Grammarly in order to improve the grammar, readability, and fluency of the English language used. After using this tool, the authors reviewed and edited the work as needed and take full responsibility of the  publication content.

\section*{Acknowledgments}
The LES and mesoscale simulations were carried out in the Winds of the North Sea in 2050 (WINS50) project that has been sponsored by the Ministry of Economic Affairs and
Climate, National regulations EZ-subsidies, Topsector Energy executed by Netherlands Enterprise Agency (project number
TEWZ119007). The authors thank all contributors of the WINS50 project for their contributions.

AWS is acknowledged for their support in producing the data and hosting the LES data on the AWS Open Data Registry. 

\appendix

\section{Engineering wake-model configurations and formulations}
\label{app}

This appendix documents the run-specific numerical implementation and the model chains used in the engineering calculations.

\subsection{Numerical implementation}
\label{app:numerical_implementation}

All calculations used the \texttt{Iterative} algorithm in
\textsc{FOXES}. Denoting the change between successive iterations by $\Delta_{\mathrm{iter}}$, convergence was assessed using the following tolerances on the maximum absolute changes:
\begin{equation}
\begin{aligned}
\max\left|\Delta_{\mathrm{iter}}U_{\mathrm{REWS}}\right|
&\leq 10^{-6}~\mathrm{m\,s^{-1}},\\
\max\left|\Delta_{\mathrm{iter}}\mathrm{TI}\right|
&\leq 10^{-7},\\
\max\left|\Delta_{\mathrm{iter}}C_T\right|
&\leq 10^{-7}.
\end{aligned}
\label{eq:iterative_convergence}
\end{equation}
Model~A did not specify \texttt{max\_\allowbreak{}it} and therefore retained the \textsc{FOXES} default of no finite iteration limit. For Models~B and C, the execution scripts set \texttt{max\_\allowbreak{}it} from the environment variable \texttt{FOXES\_\allowbreak{}MAX\_\allowbreak{}IT}, for which the script default was 14. These configurations were therefore subject to the resulting finite iteration limit.

The exact wake-model selections were \texttt{Jensen\_\allowbreak{}linear\_\allowbreak{}k004} for Model~A; \texttt{Bastankhah\allowbreak{}2014\_\allowbreak{}linear\_\allowbreak{}ka035}
together with \texttt{Crespo\allowbreak{}Hernandez\_\allowbreak{}max\_\allowbreak{}ka035} for Model~B; and \texttt{TurbOPark\_\allowbreak{}quadratic\_\allowbreak{}loc\_\allowbreak{}ka004} together with \texttt{IECTI2019\_\allowbreak{}max} for Model~C. In these names, \texttt{k004} denotes the constant value $k=0.04$,
whereas \texttt{ka035} and \texttt{ka004} denote $k=0.35\,\mathrm{TI}$ and $k=0.04\,\mathrm{TI}$, respectively, using the local \textsc{FOXES} turbulence-intensity variable. The suffixes \texttt{linear} and \texttt{quadratic\_\allowbreak{}loc} identify the corresponding velocity-superposition rules, whereas \texttt{max} identifies the wake-added-TI superposition rule. For Model~B, the far-wake parameters of the Crespo--Hern\'andez model were replaced directly in the installed source by the values of \cite{zehtabiyan2024wind}. This modification is therefore not represented by the \texttt{Crespo\allowbreak{}Hernandez\_\allowbreak{}max\_\allowbreak{}ka035} model name alone.

Wake propagation was represented with the \texttt{Streamlines\allowbreak{}2D} wake frame. For each atmospheric state and wake-generating turbine, the streamline was initialised at the rotor centre and advanced in 128~m steps using the locally interpolated horizontal wind direction at hub height. This step length was
numerically equal to the horizontal LES-grid spacing. Evaluation
locations were expressed in terms of the downstream distance $x$
along the resulting streamline and the radial distance $r$ from it. The procedure therefore represents horizontal variations in wind direction, but does not describe a vertically varying wake trajectory.

The \textsc{FOXES} versions and associated wake-length defaults
differed among the configurations. Models~A and C were run with
versions earlier than v1.7.6, for which the default maximum length of \texttt{Streamlines\allowbreak{}2D} was 20~km. Model~B was run with v1.7.6 or later and retained \texttt{max\_\allowbreak{}wake\_\allowbreak{}length\_\allowbreak{}km=\allowbreak{}None}, which imposed no explicit maximum wake length.

All configurations used the \texttt{grid9} rotor model. The standard \textsc{FOXES} partial-wake mapping applied \texttt{top\_\allowbreak{}hat} to the Jensen velocity wake, \texttt{axiwake6} to the axisymmetric Gaussian velocity wakes and \texttt{top\_\allowbreak{}hat} to the added-TI wakes. These partial-wake models converted the wake fields into rotor-effective
contributions before the applicable velocity- or TI-superposition rule was applied.

\subsection{Notation and common operations}
\label{app:common_operations}

Throughout this appendix, $j$ denotes the wake-generating turbine, $i$ the target turbine, $p$ a rotor-sampling point and $n$ the iteration index. The quantities $D_j$ and $C_{T,j}$ are the rotor diameter and thrust coefficient of the wake-generating turbine. Its ambient and local wake-corrected turbulence intensities are denoted by $\mathrm{TI}_{0,j}$ and $\mathrm{TI}_{\mathrm{eff},j}$, respectively; the corresponding quantities at the target turbine are $\mathrm{TI}_{0,i}$ and $\mathrm{TI}_{\mathrm{eff},i}$. The symbol $a_j$ denotes the axial induction factor used by the velocity-deficit model; its relation to $C_{T,j}$ is model-dependent and is specified separately for Models~A--C.

The dimensionless velocity deficit is defined as negative, such that a single-wake contribution satisfies $\delta_j<0$. The symbol $\widehat{\delta}_{j\rightarrow i}$ denotes the rotor-effective velocity-deficit contribution of turbine $j$ at turbine $i$ after application of the relevant partial-wake model. Similarly, $\widehat{\Delta\mathrm{TI}}_{j\rightarrow i}\geq0$ denotes a rotor-effective wake-added-TI contribution. These hatted quantities already include the effect of partial wake intersection and therefore require no additional geometric-overlap factor. Superscripts denoting the iteration are omitted from the individual-wake equations and retained where needed in the farm-level superposition equations.

Let $\mathcal{W}_i$ denote the set of wake-generating turbines whose wakes affect target turbine $i$. For Models~B and C, the local wake-corrected turbulence intensity $\mathrm{TI}_{\mathrm{eff},i}$ is obtained by taking the largest rotor-effective wake-added-TI contribution from this set and combining it quadratically with the ambient value $\mathrm{TI}_{0,i}$:
\begin{equation}
\mathrm{TI}_{\mathrm{eff},i}^{(n+1)}
=
\left[
\mathrm{TI}_{0,i}^2
+
\left(
\max_{j\in\mathcal{W}_i}
\widehat{\Delta\mathrm{TI}}_{j\rightarrow i}^{(n)}
\right)^2
\right]^{1/2}.
\label{eq:ti_max}
\end{equation}

For rotor $i$, let $U_{i,p}$ and $U_{0,i,p}$ denote the waked and
ambient wind speeds, respectively, at rotor-sampling point $p$, and let $\omega_p$ be the corresponding normalised quadrature weight. The waked and ambient rotor-equivalent wind speeds are denoted by $U_{\mathrm{REWS},i}$ and $U_{0,i}^{\mathrm{REWS}}$, respectively. The \texttt{grid9} rotor model calculates them as
\begin{equation}
\begin{gathered}
U_{\mathrm{REWS},i}
=
\sum_{p=1}^{9}\omega_p U_{i,p},\\
U_{0,i}^{\mathrm{REWS}}
=
\sum_{p=1}^{9}\omega_p U_{0,i,p},
\qquad
\sum_{p=1}^{9}\omega_p=1.
\end{gathered}
\label{eq:rews_grid9}
\end{equation}
Other rotor-equivalent variables used internally by the turbine type, such as those entering thrust or power calculations, retain their definitions in the corresponding \textsc{FOXES} turbine model.

\subsection{Model A: Jensen wake model}

Model~A employs the Jensen top-hat wake formulation \cite{jensen1983note}. For a downstream distance $x$ from a rotor of diameter $D_j$, the dimensionless velocity deficit at radial distance $r$ from the wake centreline is expressed in terms of the wake radius $R_{w,j}(x)$ as
\begin{equation}
\delta_j(x,r)=
\begin{cases}
-2a_j\left(\dfrac{D_j/2}{R_{w,j}(x)}\right)^2,
    & r\leq R_{w,j}(x),\\[2mm]
0,  & r>R_{w,j}(x).
\end{cases}
\label{eq:jensen_def}
\end{equation}
For Model~A, the axial induction factor $a_j$ is obtained from the Betz actuator-disc relation,
\begin{equation}
a_j
=
\frac{1}{2}\left(1-\sqrt{1-C_{T,j}}\right).
\label{eq:betz_a}
\end{equation}
The wake-growth parameter $k_j$ is set to 0.04, so the wake radius increases linearly according to
\begin{equation}
R_{w,j}(x)
=
\frac{D_j}{2}+k_jx,
\qquad
k_j=0.04.
\label{eq:jensen_width}
\end{equation}

The rotor-effective deficits from turbines in $\mathcal{W}_i$ are
combined using the \textsc{FOXES} linear wind-speed superposition. Here $U_{\mathrm{REWS},j}^{(n)}$ is the waked rotor-equivalent wind speed at wake-generating turbine $j$ in iteration $n$:
\begin{equation}
\begin{aligned}
U_{\mathrm{REWS},i}^{(n+1)}
&=
U_{0,i}^{\mathrm{REWS}}
\\
&\quad+
\sum_{j\in\mathcal{W}_i}
U_{\mathrm{REWS},j}^{(n)}
\widehat{\delta}_{j\rightarrow i}^{(n)}.
\end{aligned}
\label{eq:wslinear_iter}
\end{equation}
Model~A contains no wake-added-TI model; its local $\mathrm{TI}$
therefore remains equal to the LES-derived background $\mathrm{TI}_0$.

\subsection{Model B: Gaussian wake with modified added turbulence}

Model~B employs the axisymmetric Gaussian velocity-deficit formulation of \cite{bastankhah2014new}. Let $\sigma_j(x)$ denote the Gaussian wake width and $a_{\mathrm{eff},j}$ the induction factor evaluated from the effective thrust coefficient. The dimensionless deficit is
\begin{equation}
\delta_j(x,r)
=
-2a_{\mathrm{eff},j}
\exp\!\left[-\frac{r^2}{2\sigma_j^2(x)}\right].
\label{eq:gauss_def}
\end{equation}
Writing $f$ for the Madsen polynomial that maps thrust coefficient to axial induction, the relation used for the Gaussian velocity wakes is
\begin{equation}
\begin{aligned}
a_j=f(C_{T,j})
&=
0.2460\,C_{T,j}
\\
&\quad+
0.0586\,C_{T,j}^2
+
0.0883\,C_{T,j}^3,
\end{aligned}
\label{eq:madsen_a}
\end{equation}
which gives values similar to those obtained from Eq.~\eqref{eq:betz_a} for $C_T<0.8$.

The effective thrust coefficient $C_{T,\mathrm{eff},j}$ and its
corresponding induction factor $a_{\mathrm{eff},j}$ are
\begin{equation}
\begin{aligned}
C_{T,\mathrm{eff},j}
&=
\frac{C_{T,j}}{8\left[\sigma_j(x)/D_j\right]^2},
\\
a_{\mathrm{eff},j}
&=
f(C_{T,\mathrm{eff},j}).
\end{aligned}
\label{eq:cteff}
\end{equation}
The numerical bounds applied internally by \textsc{FOXES} to prevent non-physical amplitudes were retained in the calculations.

The wake width is
\begin{equation}
\sigma_j(x)
=
k_jx
+
0.2\,D_j
\sqrt{
\max\!\left(
\frac{1-a_j}{1-2a_j},0
\right)
},
\label{eq:gauss_width}
\end{equation}
with the turbulence-dependent growth parameter
\cite{carbajo2018wind}
\begin{equation}
k_j
=
0.35\,\mathrm{TI}_{\mathrm{eff},j}.
\label{eq:k_modelb}
\end{equation}

For wake-added turbulence, the standard \textsc{FOXES} near-wake
branch of the Crespo--Hern\'andez formulation was retained, whereas its far-wake parameters were replaced by the turbulence-optimised values of \cite{zehtabiyan2024wind}. Consequently, Model~B does not use the original Crespo--Hern\'andez far-wake parameters \cite{crespo1996turbulence}. Here $\Delta\mathrm{TI}_j(x)$ denotes the turbulence-intensity increment generated by turbine $j$ at downstream distance $x$,
$a_{\mathrm{TI},j}=\tfrac{1}{2}(1-\sqrt{1-C_{T,j}})$ is the Betz
induction factor used by the added-TI model and $x_{\mathrm{nw},j}$ is
the near-to-far-wake transition distance. The implemented correlation is
\begingroup
\small
\begin{equation}
\Delta\mathrm{TI}_j(x)
=
\begin{cases}
0.724\,a_{\mathrm{TI},j},
& x<x_{\mathrm{nw},j},
\\[2mm]
\begin{aligned}
&0.9\,a_{\mathrm{TI},j}^{0.8325}
\mathrm{TI}_{\mathrm{eff},j}^{-0.0325}
\\
&\quad{}\times
\left(\dfrac{x}{D_j}\right)^{-0.56},
\end{aligned}
& x\geq x_{\mathrm{nw},j}.
\end{cases}
\label{eq:crespo_modified}
\end{equation}
\endgroup
The distinct symbol $a_{\mathrm{TI},j}$ distinguishes the Betz
induction factor used by the added-TI model from the Madsen induction factor $a_j$ used by the Gaussian velocity-deficit model. The maximum rotor-effective wake-added-TI contribution is combined quadratically with the ambient turbulence intensity according to Eq.~\eqref{eq:ti_max}. Because the retained setting
\texttt{use\_\allowbreak{}ambti=\allowbreak{}False} makes the Crespo--Hern\'andez correlation use the local \textsc{FOXES} turbulence-intensity variable, the wake-corrected value at the wake-generating turbine is used instead of the ambient value $\mathrm{TI}_{0,j}$. Thus, the far-wake parameters
follow \cite{zehtabiyan2024wind}, whereas the use of local turbulence intensity follows the retained \textsc{FOXES} convention.

When the near-wake distance is not prescribed separately,
\textsc{FOXES} obtains it by matching the two branches in
Eq.~\eqref{eq:crespo_modified}, giving
\begin{equation}
\frac{x_{\mathrm{nw},j}}{D_j}
=
\left[
\frac{0.724}{0.9}
a_{\mathrm{TI},j}^{0.1675}
\mathrm{TI}_{\mathrm{eff},j}^{0.0325}
\right]^{-1.7857}.
\label{eq:xnw_modified}
\end{equation}

The increment in Eq.~\eqref{eq:crespo_modified} is applied uniformly within a top-hat TI wake and is zero outside it. Let
$R_{\mathrm{TI},j}^{(B)}(x)$ denote the radius of this wake. With the configured growth parameter $k_j=0.35\,\mathrm{TI}_{\mathrm{eff},j}$,
the radius is
\begin{equation}
\begin{gathered}
\begin{aligned}
R_{\mathrm{TI},j}^{(B)}(x)
&=
2\Biggl[
k_jx +
0.25\,D_j
\sqrt{
\max\!\left(
\frac{1-a_{\mathrm{TI},j}}{1-2a_{\mathrm{TI},j}},0
\right)
}
\Biggr],
\end{aligned}
\\[1mm]
k_j=0.35\,\mathrm{TI}_{\mathrm{eff},j}.
\end{gathered}
\label{eq:crespo_radius}
\end{equation}

Models~A and B use the same linear wind-speed superposition given by Eq.~\eqref{eq:wslinear_iter}.

\subsection{Model C: TurbOPark wake model}

Model~C uses the Gaussian deficit form in Eq.~\eqref{eq:gauss_def} and the Madsen induction relation in Eqs.~\eqref{eq:madsen_a} and \eqref{eq:cteff}, but replaces the linear Gaussian-width law with the TurbOPark formulation \cite{pedersen2022turbulence}. Its shape parameters $\alpha_j$ and $\beta_j$ depend on the ambient turbulence intensity, $\epsilon_j$ is the initial-width parameter and $k_j$ is the wake-growth parameter. They are defined by
\begin{equation}
\begin{gathered}
\alpha_j=1.5\,\mathrm{TI}_{0,j},
\qquad
\beta_j=
\frac{0.8\,\mathrm{TI}_{0,j}}{\sqrt{C_{T,j}}},
\\[1mm]
\epsilon_j=
0.25
\sqrt{
\max\!\left(
\frac{1-a_j}{1-2a_j},0
\right)
},
\\[1mm]
k_j=0.04\,\mathrm{TI}_{\mathrm{eff},j}.
\end{gathered}
\label{eq:turbopark_params}
\end{equation}
In this formulation, the ambient turbulence intensity
$\mathrm{TI}_{0,j}$ enters $\alpha_j$ and $\beta_j$, whereas the
configured wake-growth parameter $k_j$ uses the local effective
turbulence intensity $\mathrm{TI}_{\mathrm{eff},j}$.

With these parameters, the Gaussian wake width is
\begin{equation}
\begin{aligned}
\sigma_j(x)
&=
D_j\Bigg[
\epsilon_j
+
\frac{k_j}{\beta_j}
\Bigg(
\sqrt{
\left(
\alpha_j+\beta_j\frac{x}{D_j}
\right)^2+1
}
-
\sqrt{1+\alpha_j^2}
\\
&\quad
-
\ln\!\left(
\frac{
\left[
\sqrt{
\left(
\alpha_j+\beta_j x/D_j
\right)^2+1
}
+1
\right]\alpha_j
}{
\left(
\sqrt{1+\alpha_j^2}+1
\right)
\left(
\alpha_j+\beta_j x/D_j
\right)
}
\right)
\Bigg)
\Bigg].
\end{aligned}
\label{eq:turbopark_width}
\end{equation}

Wake-added turbulence follows the Frandsen expression \cite{frandsen2007turbulence} adopted in IEC~61400-1:2019
\cite{IEC61400-1_2019},
\begin{equation}
\Delta\mathrm{TI}_j(x)
=
\frac{1}{
1.5+0.8\,x/(D_j\sqrt{C_{T,j}})
}.
\label{eq:frandsen_ti}
\end{equation}
The model is applied as a top-hat TI wake. Denoting its full opening angle by $\theta_{\mathrm{TI}}$, the \textsc{FOXES} value
$\theta_{\mathrm{TI}}=21.6^\circ$ gives the wake radius
\begin{equation}
R_{\mathrm{TI},j}^{(C)}(x)
=
x\tan\!\left(
\frac{\theta_{\mathrm{TI}}}{2}
\right)
=
x\tan(10.8^\circ).
\label{eq:iec_radius}
\end{equation}
The increment in Eq.~\eqref{eq:frandsen_ti} is uniform within this radius and zero outside it. Its rotor-effective contributions are combined with the ambient turbulence intensity using Eq.~\eqref{eq:ti_max}.

Finally, the velocity deficits are combined using the local quadratic superposition rule,
\begin{equation}
\begin{aligned}
U_{\mathrm{REWS},i}^{(n+1)}
&=
U_{0,i}^{\mathrm{REWS}}\Biggl[
1-
\left(
\sum_{j\in\mathcal{W}_i}
\left[
\widehat{\delta}_{j\rightarrow i}^{(n)}
\right]^2
\right)^{1/2}
\Biggr].
\end{aligned}
\label{eq:wsquadraticlocal_iter}
\end{equation}


\bibliographystyle{elsarticle-num-names.bst}

\bibliography{wecuuv_control}

\end{document}